\documentclass[aps,prb,preprint,superscriptaddress,showpacs,floatfix]{revtex4}
\usepackage{hyperref}
\usepackage{amsmath,amssymb,amsfonts}
\usepackage{graphicx}
\usepackage{color}
\DeclareGraphicsExtensions{.jpg,.pdf,.png,.eps}

\begin{document}
\newcommand{\CaP} {CaFe$_{2}$As$_{2}$ }
\newcommand{\CaCo} {Ca(Fe$_{1-x}$Co$_{x}$)$_{2}$As$_{2}$ }

\title{Control of magnetic, non-magnetic and superconducting states in annealed Ca(Fe$_{1-x}$Co$_{x}$)$_{2}$As$_{2}$}

\author{S.~Ran}
\affiliation{Ames Laboratory, Iowa State University, Ames, Iowa 50011, USA}
\affiliation{Department of Physics and Astronomy, Iowa State University, Ames, Iowa 50011, USA}
\author{S.~L.~Bud'ko}
\affiliation{Ames Laboratory, Iowa State University, Ames, Iowa 50011, USA}
\affiliation{Department of Physics and Astronomy, Iowa State University, Ames, Iowa 50011, USA}
\author{W.~E.~Straszheim}
\affiliation{Ames Laboratory, Iowa State University, Ames, Iowa 50011, USA}
\author{J.~Soh}
\affiliation{Ames Laboratory, Iowa State University, Ames, Iowa 50011, USA}
\affiliation{Department of Physics and Astronomy, Iowa State University, Ames, Iowa 50011, USA}
\author{M.~G.~Kim}
\affiliation{Ames Laboratory, Iowa State University, Ames, Iowa 50011, USA}
\affiliation{Department of Physics and Astronomy, Iowa State University, Ames, Iowa 50011, USA}
\author{A.~Kreyssig}
\affiliation{Ames Laboratory, Iowa State University, Ames, Iowa 50011, USA}
\affiliation{Department of Physics and Astronomy, Iowa State University, Ames, Iowa 50011, USA}
\author{A.~I.~Goldman}
\affiliation{Ames Laboratory, Iowa State University, Ames, Iowa 50011, USA}
\affiliation{Department of Physics and Astronomy, Iowa State University, Ames, Iowa 50011, USA}
\author{P.~C.~Canfield}
\affiliation{Ames Laboratory, Iowa State University, Ames, Iowa 50011, USA}
\affiliation{Department of Physics and Astronomy, Iowa State University, Ames, Iowa 50011, USA}
\date{\today}

\begin{abstract}

We have grown single crystal samples of Co substituted \CaP using an FeAs flux and systematically studied the effects of annealing/quenching temperature on the physical properties of these samples. Whereas the as-grown samples (quenched from 960$^\circ$C) all enter the collapsed tetragonal phase upon cooling, annealing/quenching temperatures between 350$^\circ$C and 800$^\circ$C can be used to tune the system to low temperature antiferromagnetic/orthorhomic or superconducting states as well. The progression of the transition temperature versus annealing/quenching temperature ({\itshape T}-{\itshape T}$_{anneal}$) phase diagrams with increasing Co concentration shows that, by substituting Co, the antiferromagnetic/orthorhombic and the collapsed tetragonal phase lines are separated and bulk superconductivity is revealed. We established a 3D phase diagram with Co concentration and annealing/quenching temperature as two independent control parameters. At ambient pressure, for modest {\itshape x} and {\itshape T}$_{anneal}$ values, the \CaCo system offers ready access to the salient low temperature states associated with Fe-based superconductors: antiferromagnetic/orthorhombic, superconducting, and non-magnetic/collapsed tetragonal.

\end{abstract}

\maketitle

\section{Introduction}

Since the discovery of Fe-based superconductors, FeAs-based compounds have been extensively studied.\cite{Hosono08,Rotter08,JPSJFeAs,Hosono09,Chu09,Prozorov10,Canfield10,Johnston10,Stewart11,Ni11review} Part of the reason for the extensive studies is the close proximity of the superconductivity to the antiferromagnetic and structural transitions observed in members of this family which is thought to be a key ingredient for high-{\itshape T}$_{c}$ superconductivity. Among the Fe-based superconductors, the AFe$_{2}$As$_{2}$ compounds (A = Ba\cite{Rotter08,Rotter08Ba,Ni08BaK}, Sr\cite{Yan08Sr}, Ca\cite{Ni08Ca}, members of the family called 122 because of their chemical formula) are the most extensively studied and have become model systems for understanding high-{\itshape T}$_{c}$ superconductivity in Fe-based superconductors because (in part) large, high-quality, homogeneous single crystals can be readily grown. The parent compounds of the 122 family do not manifest superconductivity at ambient pressure but rather undergo a phase transition (or tightly spaced cascade of transitions) from a high temperature tetragonal, paramagnetic state to a low temperature orthorhombic, antiferromagnetic state. Using external control parameters, such as chemical substitution\cite{Rotter08,Sefat08,Ni08BaCo,Kasahara10,Ni10TM,Thaler10,Hu11Sr} or pressure\cite{Colombier09,Matsubayashi09,Torikachvili08Ba,Kotegawa09}, the antiferromagnetic and orthorhombic phases can be systematically suppressed (and often separated); when they are suppressed sufficiently, superconductivity can develop.

The physical properties of CaFe$_{2}$As$_{2}$, although similar to those of SrFe$_{2}$As$_{2}$ and BaFe$_{2}$As$_{2}$ in many aspects, are exceptional in several ways.\cite{Canfield09Ca} First, the magnetic and structural phase transitions are strongly coupled and first order, with hysteresis of several degrees as seen in thermodynamic, transport, and microscopic measurements.\cite{Ni08Ca,Goldman08} Second, \CaP is the most pressure sensitive of the AFe$_{2}$As$_{2}$ and 1111 compounds with its magnetic/structural phase transition temperature ({\itshape T}$_{N}$/{\itshape T}$_{S}$) being initially suppressed by over 100~K per GPa but remaining sharply first order (in hydrostatic medium) as it is suppressed.\cite{Torikachvili08Ca,Kreyssig08,Goldman09,Yu09,Lee08,Prokes10,Park08} As pressure increases, a nonmagnetic, collapsed tetragonal phase that is stabilized by $\sim$0.3 GPa, intersects and terminates the lower-pressure antiferromagnetic/orthorhombic phase line near 100~K and 0.4~GPa, and rises to 300~K by $\sim$1.5~GPa.\cite{Kreyssig08,Goldman09} Therefore, there are three ground states (antiferromagnetic/orthorhombic, superconducting, and non-magnetic/collapsed tetragonal) competing at low temperature. Subsequently, the collapsed tetragonal phase was also observed in Ba122 and Sr122 under much higher pressure. (At room temperature, the pressures needed to stabilize the collapsed tetragonal in Ba122 and Sr122 are 22~GPa and 10~GPa, respectively.\cite{Uhoya10,Mittal11,Uhoya11}) Third, the physical properties of the single crystals of \CaP are remarkably dependent on the crystal growth procedure. Our previous work\cite{Ran11} has shown that crystals grown out of an FeAs flux, quenched from high temperature, exhibit a transition from the paramagnetic, tetragonal phase to the non-magnetic, collapsed tetragonal phase below 100~K at ambient pressure, in contrast to the behavior of \CaP grown from Sn flux.\cite{Ni08Ca} Further, we discovered that for the FeAs flux grown samples, a process of annealing and quenching can be used as an additional control parameter which can tune the ground state of \CaP systematically. The effects of annealing and quenching are similar to those of the pressure (as is suggested by the similarity between the annealing phase diagram and pressure phase diagram\cite{Ran11}), and this can be explained by our TEM results\cite{Ran11}, which reveal nano-scale precipitates with overlapping strain fields. It is very likely that the annealing and quenching process controls the amount of strain built up in the samples and, as a result, mimics the modest pressures needed to stabilize the collapsed tetragonal phase.

Chemical substitution, such as Co substitution, as a control parameter, has been studied extensively for members of the 122 family. For Ba122, Co substitution first suppresses the antiferromagtic/orthorhombic state and then induces superconductivity, making Ba(Fe$_{1-x}$Co$_{x}$)$_{2}$As$_{2}$ a model system for the study of high-{\itshape T}$_{c}$ superconductivity in Fe-based superconductors.\cite{Canfield10,Ni11review,Ni08BaCo,Sefat08} For Ca122, the effects of Co substitution have been studied only on the samples grown out of Sn, which have issues with solubility, reproducibility and inhomogeneity.\cite{Harnagea11,Matusiak10,Hu11Ca} The phase diagrams constructed by different groups do not match very well and therefore, need to be clarified. In this work, we studied Co substituted Ca122 grown out of an FeAs flux and by systematically control annealing/quenching temperatures we have minimized these problems. Indeed, as in the case of unsubstituted Ca122 grown out of an FeAs flux, we found that annealing/quenching temperature is a vital parameter to control and understand this system. In this paper, we present a systematic study of the combined effects of Co substitution and annealing/quenching on the physical properties of Ca122 and construct phase diagrams for different substitution levels and different annealing/quenching temperatures. Also, by combining the two control parameters, we are able to extend the 2-dimentional, {\itshape T}-{\itshape x} and {\itshape T}-{\itshape T}$_{anneal}$ ({\itshape T}$_{anneal}$ is the annealing/quenching temperature) phase diagrams into a 3-dimentional, {\itshape T}-{\itshape x}-{\itshape T}$_{anneal}$ phase diagram and reveal richer physics and better control of the system, all at ambient pressure.

\section{Experimental methods}

Single crystals of \CaCo were grown out of an FeAs flux, using conventional high-temperature solution growth techniques.\cite{Fisk89,Canfield92,Canfield01,Canfieldbook} Small Ca chunks, FeAs powder, and
CoAs powder were mixed together according to the ratio Ca:FeAs:CoAs =1:4(1-{\itshape x}$_{nominal}$):4{\itshape x}$_{nominal}$, where {\itshape x}$_{nominal}$ is the nominal Co concentration. The maximum relative error bar in {\itshape x}$_{nominal}$, as determined from potential weighing error for the lowest Co substitution level, is roughly 1.5$\%$. Single crystals were grown by rapidly cooling the Ca-Fe-Co-As melt from 1180$^\circ$C to 1020$^\circ$C over 3 h, slowly cooling from 1020$^\circ$C to 960$^\circ$C over 35 h, and then decanting off the excess liquid flux. These samples will be referred to as \textquotedblleft as-grown$\textquotedblright$ samples. In the process of decanting off the excess flux, the samples were essentially quenched from 960$^\circ$C to room temperature, which, according to our previous study,\cite{Ran11} causes strain inside the samples due to the formation of nano-precipitates of FeAs$_{x}$, leading to behavior different from Sn grown samples.\cite{Ni08Ca} Postgrowth thermal treatments of samples involve annealing samples at temperatures ranging from 350$^\circ$C to 800$^\circ$C and subsequently quenching them from this annealing temperature to room temperature. The initial determination of the {\itshape T}-{\itshape x}-{\itshape T}$_{anneal}$ phase diagrams was done by annealing/quenching individual crystals that have been picked from a growth and resealed in evacuated silica tubes. The resealed individual samples were annealed for 24 hours at annealing temperature of 400$^\circ$C or above. A longer time anneal (5 days) was used at annealing temperature of 350$^\circ$C. Once the {\itshape T}-{\itshape x}-{\itshape T}$_{anneal}$ phase diagrams were established, whole, unopened batches of samples were annealed and quenched from 350$^\circ$C or above. After 14 days at {\itshape T}$_{anneal}$ = 350$^\circ$C or after 5 days at {\itshape T}$_{anneal}$ above 350$^\circ$C, the data collected on samples from these \textquotedblleft whole batch anneals$\textquotedblright$ were quantitatively similar to those collected on the individually annealed samples. Details of the annealing and quenching technique, as well as a study of the salient annealing time scale, can be found in the previous paper.\cite{Ran11} 

Elemental analysis was performed on each these batches using wavelength-dispersive x-ray spectroscopy (WDS) in the electron probe microanalyzer of a JEOL JXA-8200 electron microprobe. Since the properties of a given sample are found to be determined by both the Co substitution level ({\itshape x}$_{WDS}$) and the post growth annealing/quenching temperature, samples will be identified by both of these parameters. For example, specific heat data will be presented for an {\itshape x} = 0.033/{\itshape T}$_{anneal}$ = 350$^\circ$C sample.

Diffraction from the plate-like samples was first performed at room temperature using a Rigaku Miniflex diffractometer with Cu {\itshape K}$\alpha$ radiation. Only (00l) peaks are observed, from which the values of the {\itshape c}-lattice parameter are inferred. Standard powder x-ray diffraction was not attempted since we have found that \CaP based compounds are very easily damaged by attempts to grind them. Diffraction lines broaden dramatically even compared to the Ba122 and Sr122.\cite{Ni08Ca} Of equal concern, the magnetization data from powder is dramatically different from that of intact crystals, probably due to gross deformation or partial amorphization during the process of \textquotedblleft grinding$\textquotedblright$ the samples. 

A temperature dependent, high resolution, single crystal x-ray diffraction measurements were performed on a representative sample using a four-circle diffractometer and Cu {\itshape K}$\alpha_{I}$ radiation from a rotating-anode x-ray source, selected by a germanium (1 1 1) monochromator. For this measurement, a sample with a dimension of 4 $\times$ 3 $\times$ 0.5~mm$^3$ was attached to a flat copper sample holder on the cold finger of a closed-cycle displex refrigerator. The mosaicity of the sample was less than 0.02$^\circ$, full width at half maximum (FWHM), as measured by the rocking curves through the (0 0 10) reflection at room temperature. The diffraction data were obtained as a function of temperature between room temperature and 6~K, the base temperature of the refrigerator. The (0 0 10) and (1 1 10) reflections were measured at each temperature and the diffraction angles were obtained from $\theta$-2$\theta$ scans in order to calculate the lattice parameters {\itshape a} and {\itshape c}. 

\begin{figure}[!htbp]
\begin{center}
\includegraphics[angle=0,width=100mm]{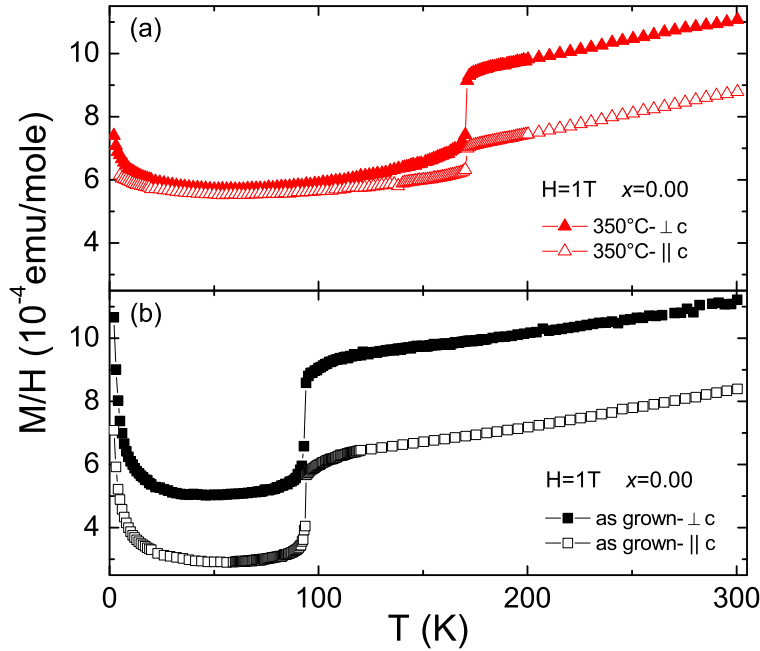}
\end{center}
\caption{(Color online) Temperature dependent anisotropic magnetic susceptibility, with a magnetic field of 1~T applied perpendicular and parrallel to the {\itshape c}-axis, for (a) the {\itshape x} = 0.00/{\itshape T}$_{anneal}$ = 350$^\circ$C sample, as an example of the antiferromagnetic/structural transition, and (b) the {\itshape x} = 0.00/as-grown sample, as an example of the collapsed tetragonal phase transiton.}
\label{anisotropy}
\end{figure}

Temperature dependent magnetization measurements were made in a Quantum Design Magnetic Property Measurement System (MPMS). It turns out that when the magnetic field is applied parallel to the {\itshape c}-axis, the size of the jump in the magnetic susceptibility for the collapsed tetragonal phase transition is significantly larger than that for the antiferromagnetic/structural phase transition, whereas, when the magnetic field is applied perpendicular to the {\itshape c}-axis, the two types of transitions manifest comparable sized jumps in magnetic susceptibility (Fig. \ref{anisotropy}). For low annealing/quenching temperatures, only the antiferromagnetic/structural phase transition exists for all Co concentration in our study and susceptibility was measured with the applied magnetic field perpendicular to the {\itshape c}-axis. On the other hand, for higher annealing/quenching temperature, the collapsed tetragonal phase transition occurs for higher Co concentration. Therefore susceptibility was measured with applied magnetic field parallel to the {\itshape c}-axis in order to allow for clearer differentiation between the two types of transition.

The in plane, temperature dependent electrical AC ({\itshape f} = 16~Hz, {\itshape I} = 1~mA) resistance measurements were performed in Quantum Design MPMS systems operated in external device control (EDC) mode, in conjunction with Linear Research LR700 AC resistance bridges. The electrical contacts were placed on the samples in standard 4-probe geometry, using Pt wires attached to the sample with Epotek H20E silver epoxy. The temperature dependent AC ({\itshape f} = 16~Hz, {\itshape I} = 1~mA) resistance was also measured in applied magnetic field up to 14~T in a Quantum Design Physical Property Measurement System (PPMS) so as to determine the anisotropic, upper superconducting critical field, {\itshape H}$_{c2}$({\itshape T}) values. Temperature dependent heat capacity for representative samples was measured in Quantum Design PPMS systems using the relaxation technique in both zero field and magnetic field of either 9~T or 14~T applied along the {\itshape c}-axis.

\begin{figure}[!htbp]
\begin{center}
\includegraphics[angle=0,width=100mm]{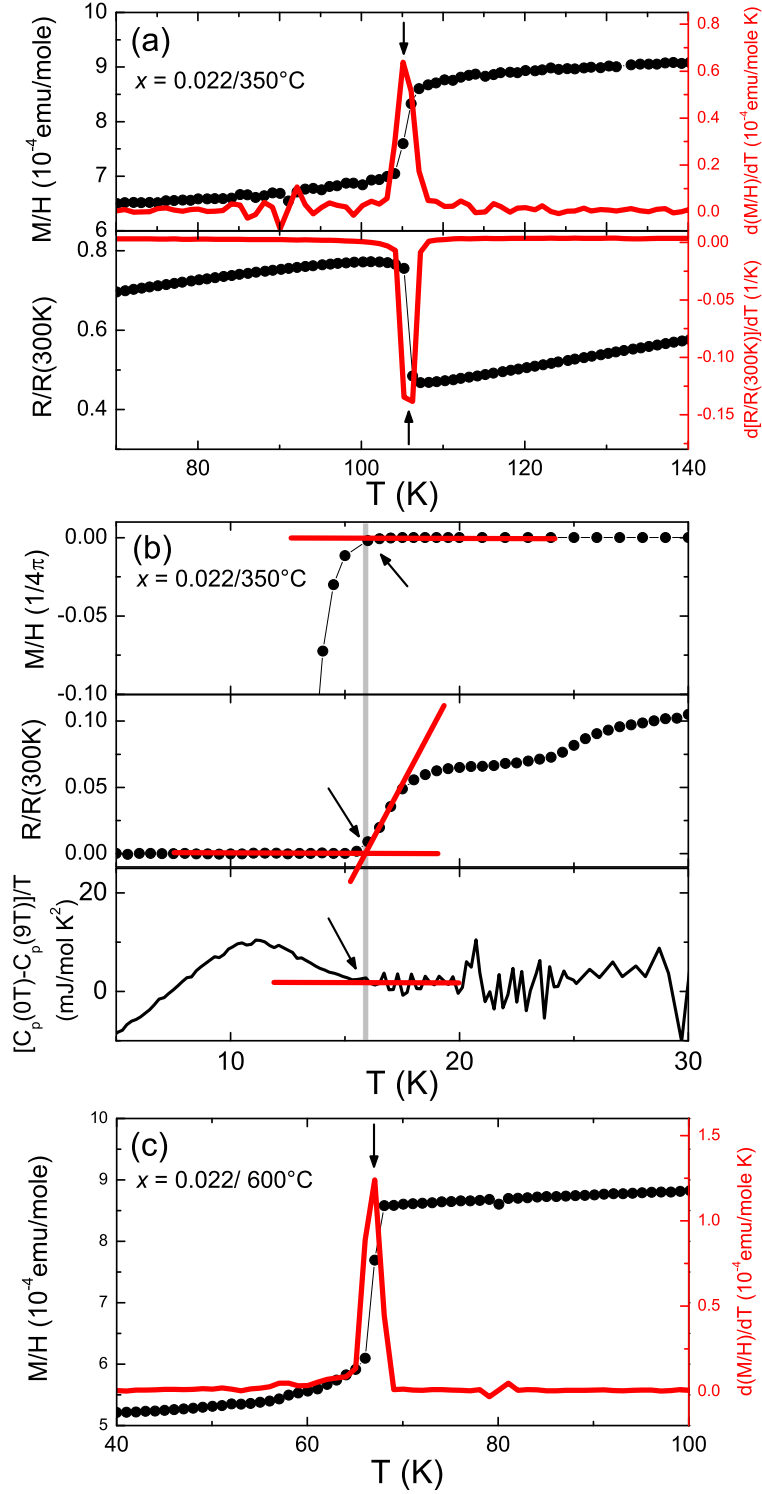}
\end{center}
\caption{(Color online) Criteria used to determine the transition temperatures of (a) the antiferromagnetic/structural phase transition, (b) the superconducting phase transition and (c) the collapsed tetragonal phase transition. Inferred transition temperature is indicated by arrow in each figure.}
\label{criteria}
\end{figure}

In order to infer phase diagrams from these thermodynamic and transport data, we need to introduce criterion for determination of the salient transition temperatures. For almost all combinations of Co concentration and annealing/quenching temperature, the antiferromagnetic/structural phase transition (when present) appears as a single sharp feature which is clearly identifiable in both resistance and magnetization. Figure \ref{criteria}a shows the susceptibility and resistance, as well as their temperature derivatives, for a {\itshape x} = 0.022/{\itshape T}$_{anneal}$ = 350$^\circ$C sample. Clear features, including a sharp drop in susceptibility and a sharp jump in the resistance, occur upon cooling through the transition temperature. The transition temperature is even more clearly seen in the $d(M/H)/dT$ and $dR/dT$ data. In a similar manner criteria for the determination of {\itshape T}$_{c}$ have to be established and used. For this study we use an onset criterion for susceptibility (the temperature at which the susceptibility deviates from the normal-state susceptibility) and an offset criterion for resistance (the temperature at which the maximum slope of the resistance data that goes to zero resistance extrapolates to zero resistance). The criteria for {\itshape T}$_{c}$ are presented in Fig. \ref{criteria}b, with an example of a {\itshape x} = 0.033/{\itshape T}$_{anneal}$ = 350$^\circ$C sample. For comparison, specific heat data for this sample are also presented. It can be seen that the {\itshape T}$_{c}$ values inferred from both susceptibility and resistance data, as well as that from the onset criterion for specific heat (the temperature at which the specific heat deviates from the normal-state specific heat), match quite well. The collapsed tetragonal phase is induced by higher annealing/quenching temperatures. When the collapses tetragonal phase transition is first order, it often leads to cracks in the resistance bar and loss of data below the transition temperature, which makes it difficult to extract the transition temperature from {\itshape R}({\itshape T}) data. Therefore susceptibility data were primarily used to determine {\itshape T}$_{cT}$. Figure \ref{criteria}c shows the temperature derivative of the susceptibility data, with a sharp peak, which was employed to determine {\itshape T}$_{cT}$. 

\section{Results}

A summary of the WDS measurement data is presented in Table 1. The table shows the nominal concentration, the measured average {\itshape x} value, and twice the standard deviation of the {\itshape x} values measured. For each sample, the measurement was done at 12 different locations on a cleaved surface. Data points of nominal versus actual concentration can be fit very well with a straight line, with a slope of 0.96 $\pm$ 0.01, indicating a linear correlation between the measured Co concentration and the nominal concentration. The error bars are taken as twice the standard deviation determined from the measurements, and the largest deviation from the nominal value is no more than 0.002, demonstrating relative homogeneity of the substituted samples studied here. In the following, the average experimentally determined {\itshape x} values, {\itshape x}$_{WDS}$, will be used to identify all the compounds rather than the nominal concentration, {\itshape x}$_{nominal}$. These results are in stark contrast to the non-monotonic and scattered {\itshape x}$_{WDS}$ versus {\itshape x}$_{nominal}$ results found for the \CaCo grown from Sn, for which solubility problems in Sn make systematic measurements on homogeneous samples difficult.\cite{Harnagea11,Matusiak10,Hu11Ca}

\begin{table}
  \caption{WDS data for Ca(Fe$_{1-x}$Co$_{x}$)$_{2}$As$_{2}$. {\itshape x}$_{nominal}$ is the nominal concentration of the substitutions. {\itshape x}$_{WDS}$ is the average of {\itshape x} values measured at 12 locations on samples from each batch. 2$\sigma$ is twice the standard deviation of the 12 values measured.}
\vspace{5mm}
{\begin{tabular}{@{}lcccccccc}
\toprule
\multicolumn{9}{c}{Ca(Fe$_{1-x}$Co$_{x}$)$_{2}$As$_{2}$ } \\ 
\colrule
   {\itshape x}$_{nominal}$ &0.01	&0.02	&0.025	&0.03	&0.035	&0.04   &0.05	&0.06\\
   {\itshape x}$_{WDS}$ &0.010	&0.019	&0.022	&0.028	&0.033	&0.038  &0.049	 &0.059\\
   2$\sigma$	&0.001	&0.001	&0.001	&0.001	&0.001	&0.001  &0.002	&0.002\\
\botrule
  \end{tabular}}

\end{table}

\begin{figure}[!htbp]
\begin{center}
\includegraphics[angle=0,width=100mm]{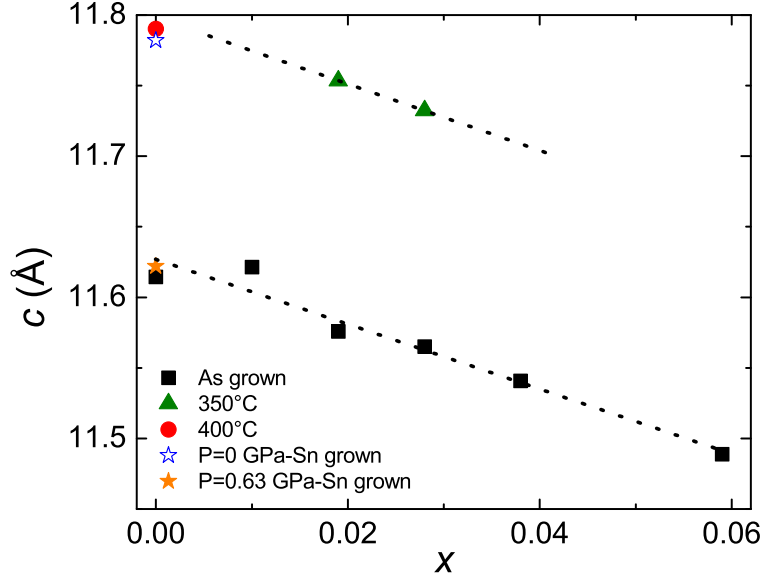}
\end{center}
\caption{(Color online) Room temperature {\itshape c}-lattice parameter of Ca(Fe$_{1-x}$Co$_{x}$)$_{2}$As$_{2}$ of as-grown samples and samples annealed/quenched at selected temperatures as a function of measured Co concentration, {\itshape x}, determined via the diffraction from plate-like samples described in the Experimental Methods section. For comparision, data of Sn grown sample under pressure are also presented.\cite{Kreyssig08,Ran11} Black dotted lines are the guide to the eyes.}
\label{clattice}
\end{figure}

Figure~\ref{clattice} presents the {\itshape c}-lattice parameters of the as-grown samples, as well as selected annealed samples, determined via the diffraction from plate-like samples described above, using the (002) and (008) peaks. The {\itshape x} = 0.00/{\itshape T}$_{anneal}$ = 400$^\circ$C sample has {\itshape c}-lattice parameter similar to that of the Sn grown sample\cite{Kreyssig08,Ran11} whereas the as-grown sample manifests a reduction of almost 2$\%$ in the {\itshape c}-lattice parameter. Data for Sn grown \CaP at ambient and applied pressure of {\itshape P} = 0.63~GPa demonstrate that the effects of applied pressure and annealing/quenching temperature are remarkably similar. (It should be noted that both the Sn grown sample under 0.63~GPa pressure and the FeAs grown sample in the as-grown state transform to the collapsed tetragonal phase upon cooling below 200~K.\cite{Kreyssig08,Ran11}) Substituting Co decreases {\itshape c}-lattice parameter for both annealed/quenched samples and as-grown samples, at roughly the same rate.	

\begin{figure}[!htbp]
\begin{center}
\includegraphics[angle=0,width=100mm]{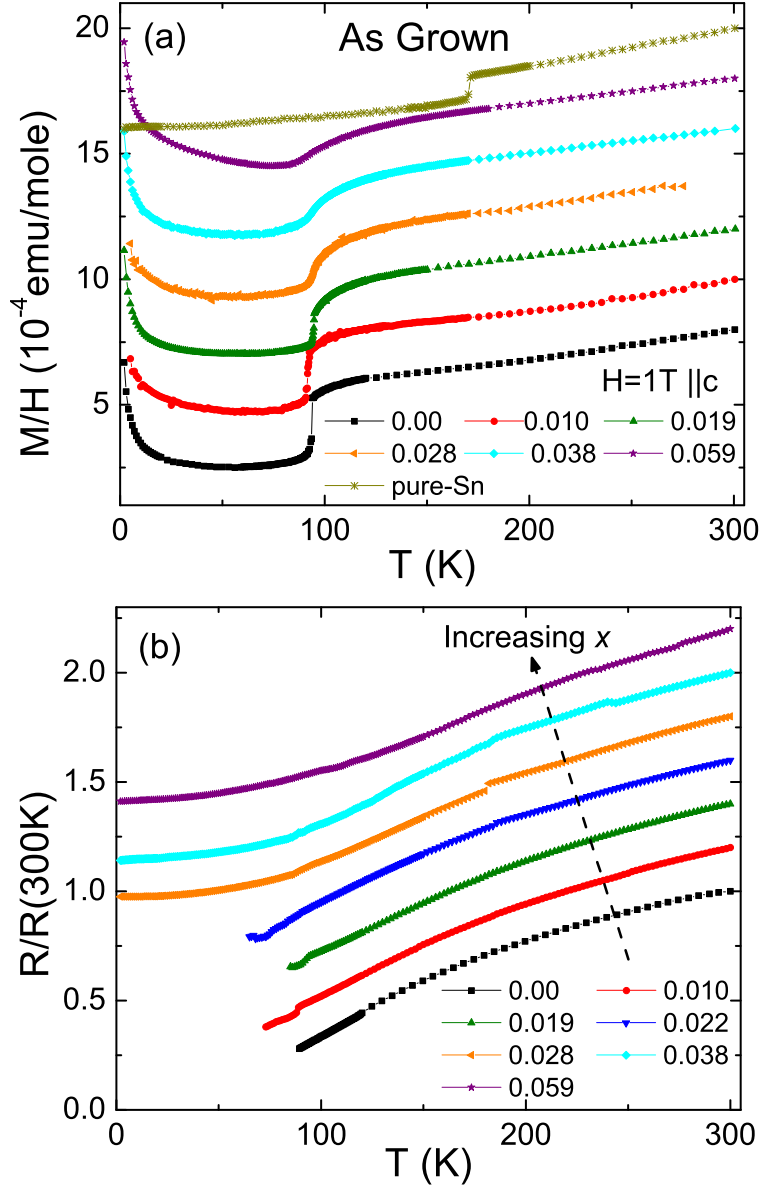}
\end{center}
\caption{(Color online) Temperature dependent (a) magnetic susceptibility with a field of 1 T applied parallel to the {\itshape c}-axis and (b) normalized electrical resistance of as-grown Ca(Fe$_{1-x}$Co$_{x}$)$_{2}$As$_{2}$ samples. For clarity, susceptibility data in (a) have been offset by 2 $\times$ 10$^{-4}$ emu/mole from each other and resistance data in (b) have been offset by 0.2 from each other.}
\label{Asgrown-MR}
\end{figure}

Figures~\ref{Asgrown-MR}a and \ref{Asgrown-MR}b present the temperature dependent magnetic susceptibility, with magnetic field applied parallel to the {\itshape c}-axis, and normalized resistance for \CaCo single crystal, as-grown, samples with Co substitution levels up to {\itshape x} = 0.059. For the pure compound, CaFe$_{2}$As$_{2}$, the susceptibility of the as-grown sample shows a sharp drop ($\sim$50$\%$) below 100~K, which is associated with a phase transition from the high temperature, tetragonal, paramagnetic state to the low temperature, collapsed tetragonal, non-magnetic state.\cite{Ran11} Note that the size of the jump is almost twice as large as that of the antiferromagnetic/structural phase transition of the Sn grown sample (top of Fig.~\ref{Asgrown-MR}a) when measured with field parallel to the {\itshape c}-axis. This phase transition can produce a downward jump in resistance when cooling down,\cite{Yu09} but, given that this is a first order, structural phase transition, it often leads to cracking along the length and width of the bar, as well as loss of contacts. For these reasons resistance data simply stops as temperature drops below {\itshape T}$_{cT}$.	

\begin{figure}[!htbp]
\begin{center}
\includegraphics[angle=0,width=100mm]{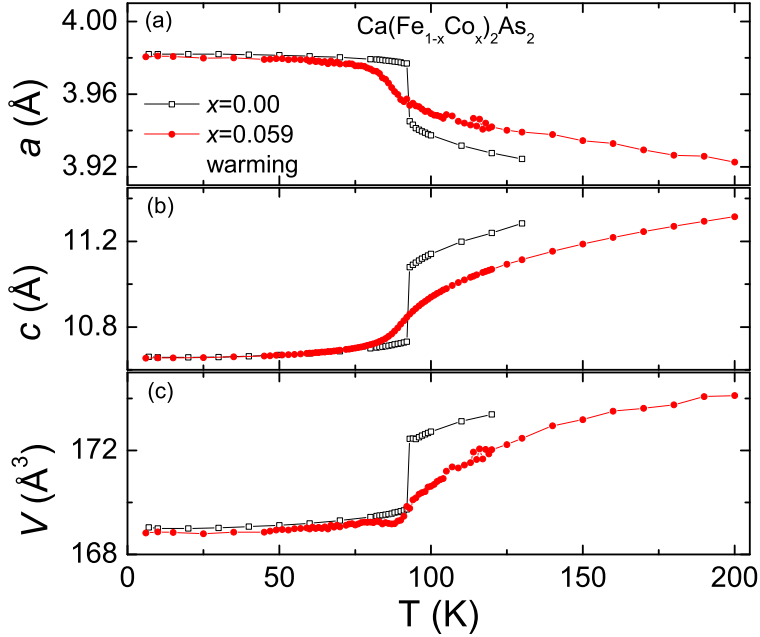}
\end{center}
\caption{(Color online) Values for (a) the {\itshape a}-lattice paramter, (b) the {\itshape c}-lattice parameter and (c) unit
cell volume as a function of temperature for \CaCo for {\itshape x} = 0.00 (square)\cite{Ran11} and {\itshape x} = 0.059 (circle) as-grown samples determined from single crystal X-ray diffraction measurements.}
\label{latticeT}
\end{figure}

For low Co substitution values, the magnetic susceptibility shows little change with the signature of the phase transition appearing at roughly the same temperature. The only significant change in the magnetization data is the loss of the discontinuos jump in {\itshape M}({\itshape T})/{\itshape H} on cooling for {\itshape x} = 0.028 and higher. In order to confirm that the low temperature state of the Co substituted, as-grown, samples is a tetragonal phase with reduced {\itshape c}-lattice parameter, a temperature dependent, single crystal x-ray measurement was carried out on the {\itshape x} = 0.059 sample. Figure~\ref{latticeT} displays the temperature dependence of the lattice parameters as well as the unit cell volume. For the {\itshape x} = 0.00 and {\itshape x} = 0.059, as-grown samples, it is clear that there is a reduction of the {\itshape c}-lattice parameter and an expansion of the {\itshape a}-lattice parameter from high temperature to low temperature. The overall unit cell volume shrinks as a result. The lattice parameters for the {\itshape x} = 0.059 sample are almost the same as those for the pure compound at low temperature. 

However, both the changes in the lattice parameters and the magnetic susceptibility of the {\itshape x} = 0.059 sample are dramatically broadened comparing with those of the pure compound. Instead of a sharp jump at the transition temperature indicating a first order phase transition, the lattice parameters and the magnetic susceptibility change gradually over $\sim$30 K. Moreover, this broadening in signatures of transition coincides with the changes in the resistance data, with the resistance bar surviving as it is cooled down to the base temperature of 1.8~K, instead of cracking and losing contact which is often an indication of a strongly first order structural phase transition. All these thermodynamic, transport and microscopic measurements suggest the possibility that a critical end point of the phase transition may exist and, at {\itshape x} = 0.059, the system has already gone beyond this critical end point resulting in a continuous thermal contraction rather than a first order phase transition. Further structural investigetions of this issue are planned.  
 
The results presented above for the as-grown \CaCo samples are dramatically different from those reported for the Sn grown samples.\cite{Harnagea11,Matusiak10,Hu11Ca} In the case of the pure compound, this difference is caused by stress and strain built up inside the sample during the process of quenching from 960$^\circ$C.\cite{Ran11} Control of post growth annealing and quenching can systematically suppress the magnetic/structural transition and stabilize the collapsed tetragonal phase in a manner analogous to applied pressure. For Ca(Fe$_{1-x}$Co$_{x}$)$_{2}$As$_{2}$, we expect the annealing/quenching temperatures to serve as a tuning parameter in a similar way. In order to study the effect of the annealing/quenching temperature on the Co substituted samples, we annealed and quenched the samples with different concentrations at temperatures ranging from 350$^\circ$C to 800$^\circ$C and measured their thermodynamic and transport properties.

\begin{figure}[!htbp]
\begin{center}
\includegraphics[angle=0,width=100mm]{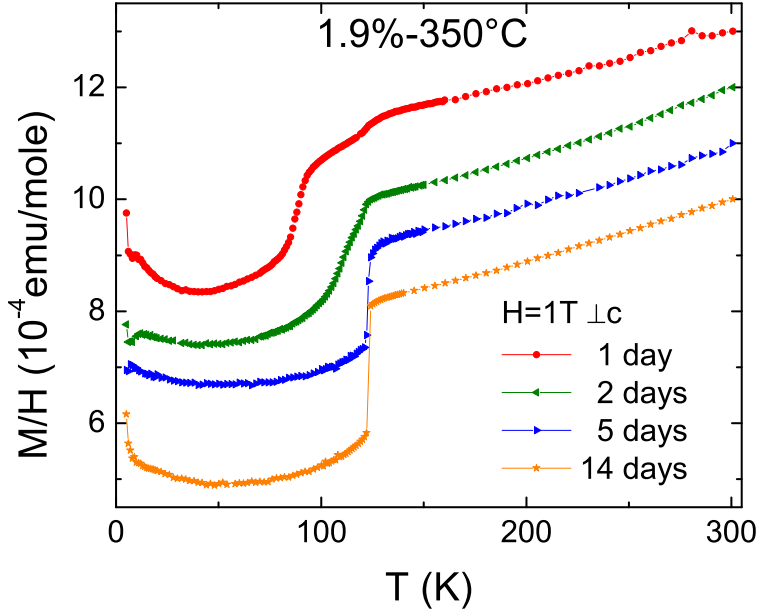}
\end{center}
\caption{(Color online) Temperature dependent magnetic susceptibility with a field of 1 T applied perpendicular to the {\itshape c}-axis of {\itshape x} = 0.019 samples annealed at 350$^\circ$C for different amount of time varing from 1 day to 14 days. Susceptibility data have been offset by 1 $\times$ 10$^{-4}$ emu/mole from each other for clarity.}
\label{1.9-350C}
\end{figure}

For annealing/quenching temperature equal or above 400$^\circ$C, samples were annealed for 24 hours. As we discussed in the previous paper,\cite{Ran11} for these temperatures, 24 hours is longer than the time needed to reach a well defined state. For the annealing/quenching temperature of 350$^\circ$C, we determined the annealing time in a similar way. In Fig.~\ref{1.9-350C} we show the evolution of the magnetic susceptibility for different annealing times at 350$^\circ$C. It is clear that 24 h is not a sufficient amount of time to reach a well defined state. It leads to split, broadened features with drops in susceptibility below both 130 and 100~K. 48~h leads to a less split but still broadened feature near 125~K. 5 days leads to a single, sharp feature at around 125~K, which is comparable to what is seen for a 14 day anneal. This progression shows that for 350$^\circ$C, the salient time scale for annealing is between 2 and 5 days. Therefore, for 350$^\circ$C, samples were annealed for 5 days. In the case where whole batches were annealed without opening, the annealing time used was longer, for example, for 350$^\circ$C, the whole batches were annealed for 14 days.

\begin{figure}[!htbp]
\begin{center}
\includegraphics[angle=0,width=160mm]{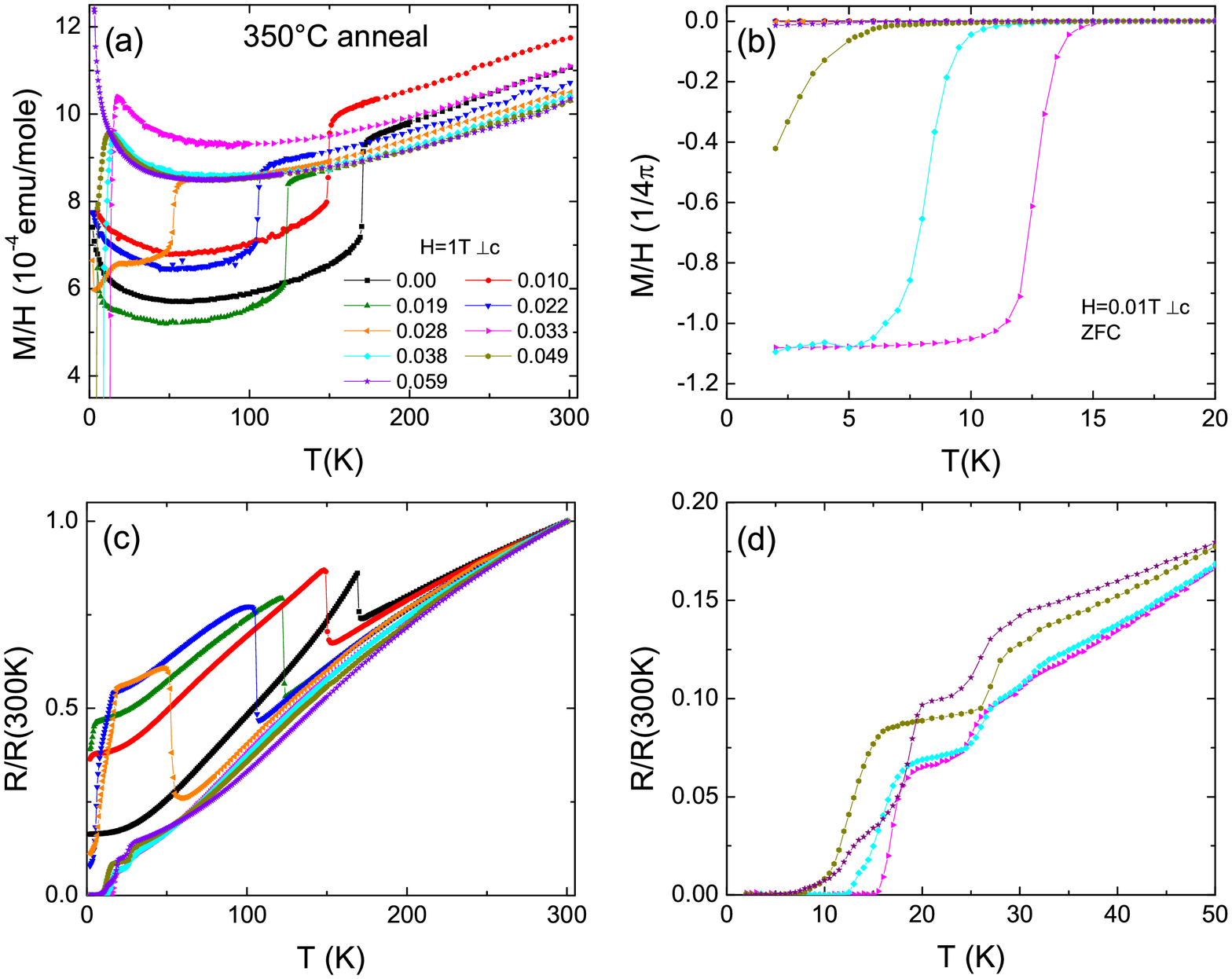}
\end{center}
\caption{(Color online) Temperature dependent (a) magnetic susceptibility with a field of 1~T applied perpendicular to the {\itshape c}-axis, (b) low field magnetic susceptibility measured upon zero field cooling (ZFC) with a field of 0.01 T applied perpendicular to the {\itshape c}-axis and (c) normalized electrical resistance of Ca(Fe$_{1-x}$Co$_{x}$)$_{2}$As$_{2}$ samples for an annealing/quenching temperature of 350$^\circ$C. Low temperature resistance of superconducting samples are presented in (d).}
\label{350CMR}
\end{figure}

Figure~\ref{350CMR}a presents the in plane susceptibility data in a field of 1~T applied perpendicular to the {\itshape c}-axis for annealing/quenching temperature of 350$^\circ$C.  After being annealed/quenched at 350$^\circ$C, the pure compound manifests a magnetic/structural phase transition at around 170~K from the high temperature, tetragonal, paramagnetic state to the low temperature, orthorhombic, antiferromagnetic state as indicated by the sharp, modest, drop in susceptibility (and sharp increase in resistance shown in Fig.~\ref{350CMR}c, as will be discussed momentarily).\cite{Ran11} This phase transition is progressively suppressed by Co substitution until it is completely suppressed by {\itshape x} = 0.033. The magnetic signature of the phase transition remains quite sharp with the size of the jump fairly constant. The superconducting phase first appears in the {\itshape x} = 0.033 sample, with the superconducting transition temperature {\itshape T}$_{c}$ around 15~K. As the Co substitution level is further increased, {\itshape T}$_{c}$ decreases. An upper limit of the superconducting fraction can be obtained from the zero field cooling susceptibility in the field of 0.01~T as shown in Fig.~\ref{350CMR}b.  Approximately 100$\%$ of diamagnetism is seen for the {\itshape x} = 0.033 and {\itshape x} = 0.038 samples without taking account of demagnetization factor. For higher Co substitution the diamagnetic fraction decreases and becomes essential zero for the {\itshape x} = 0.059 sample.

Figure~\ref{350CMR}c shows normalized, temperature dependent resistance data for the 350$^\circ$C annealed/quenched samples. For substitution levels up to {\itshape x} = 0.028, the antiferromagnetic/structural phase transition is further confirmed by the same sharp, upward jump in resistance, similar to that found in pure Ca122. As the transition temperature is suppressed, this signature remains sharp while the size of the jump increases monotonically and reaches 40$\%$ of room temperature resistance value at {\itshape x} = 0.028. The increasing size of the jump with suppressing {\itshape T}$_{N}$/{\itshape T}$_{S}$ is similar to what has been seen for the pure compound grown out of an FeAs flux, but is in contrast to the case of Sn grown samples under pressure,\cite{Yu09} where the size of the jump remains relatively constant. Although the resistance starts to decrease at low temperature for the samples with {\itshape x} = 0.019, 0.022 and 0.028, it does not reach instrumental zero. Considering that low field susceptibility does not show significant diamagnetism, the sudden drop in resistance for these three samples most likely indicates filamentary superconductivity.\cite{Saha09,Colombier09} Complete superconducting phase transitions with zero resistance are obtained for {\itshape x} $\geq$ 0.033. The fact that resistance shows several steps before reaches instrumental zero, the highest of which has an onset near 30~K, suggest that there may be some microscopic inhomogeneity of the stress and strain. This will be discussed in detail in the Discussion section below. {\itshape T}$_{c}$ decreases gradually with increasing Co concentration and drops to around 2.5~K for {\itshape x} = 0.059. Again, since the diamagnetic fraction for this concentration is essentially zero, it may be a filamentary superconductor. 

Before we proceed further, it is important to further explore whether that the superconductivity at optimal substitution and annealing/quenching temperature is a bulk property instead of filamentary superconductivity since zero resistance can be caused by only a thin layer or filament spanning the sample. Low field susceptibility, as a thermodynamic quantity, is normally used to confirm the bulk superconductor. However, the low field susceptibility was measured after cooling in a zero applied field, and therefore only tells the upper limit of the superconducting fraction. 

\begin{figure}[!htbp]
\begin{center}
\includegraphics[angle=0,width=100mm]{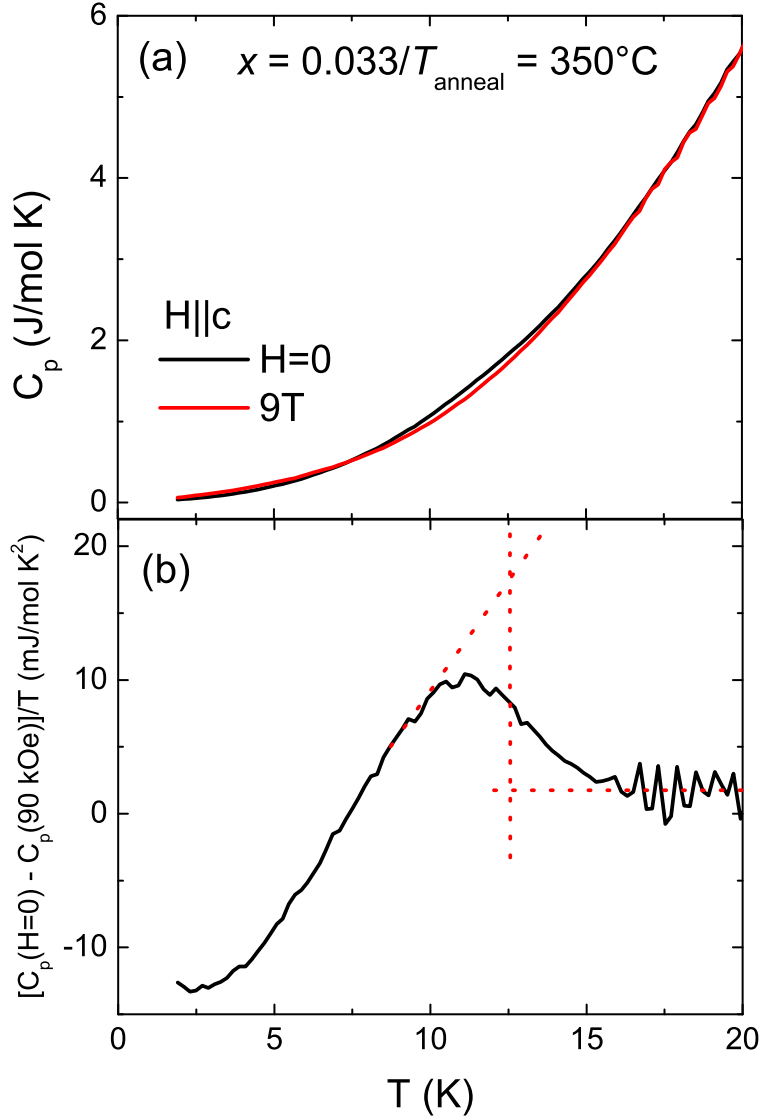}
\end{center}
\caption{(Color online) (a) Temperature dependent specific heat data of the {\itshape x} = 0.033/{\itshape T}$_{anneal}$ = 350$^\circ$C sample, measuered in zero field and a field of 9~T applied parallel to the {\itshape c}-axis and (b) the difference between of the two sets of data presented as $\Delta${\itshape C}$_{P}$/{\itshape T}. The red dased lines represent the isoentropic construction used to determine the jump in {\itshape C}$_{P}$ at {\itshape T}$_{c}$ (see text).}
\label{3.3-350C-Cp}
\end{figure}

One way to further establish that bulk superconductivity is present is to measure the temperature dependent specific heat and determine the size of the jump at {\itshape T}$_{c}$. Figure~\ref{3.3-350C-Cp} presents the specific heat data on a representative sample, {\itshape x} = 0.033/{\itshape T}$_{anneal}$ = 350$^\circ$C, which shows full diamagnetism from zero field cooled-warming susceptibility data. Specific heat was measured in both zero field and in 9~T and the size of the jump in {\itshape C}$_{P}$ at {\itshape T}$_{c}$ can be inferred from the difference between these two data sets. (As will be shown below, anisotropic {\itshape H}$_{c2}$({\itshape T}) data on an optimal substituted/annealed \CaCo samples show that 9~T is an adequate field for this substraction and analysis.) Due to finite widths of the superconducting transitions, $\Delta${\itshape C}$_{P}$/{\itshape T}$_{c}$ and {\itshape T}$_{c}$ values were determined from {\itshape C}$_{P}$/{\itshape T} vs {\itshape T} data using an “isoentropic” construction (i.e., such that the vertical line in Fig.~\ref{3.3-350C-Cp}b delineates equal areas in the {\itshape C}$_{P}$/{\itshape T} vs {\itshape T} plot). A  $\Delta${\itshape C}$_{P}$/{\itshape T}$_{c}$ value of 16.1~mJ/mol~K$^2$ is inferred from this criterion. These data fall onto a manifold of $\Delta${\itshape C}$_{P}$/{\itshape T}$_{c}$ versus {\itshape T}$_{c}^{2}$ data found for many substituted AFe$_{2}$As$_{2}$ materials\cite{Stewart11,Sergey09cp,Kogan09,Matsuda11} (see discussion below), suggesting that there is bulk superconductivity in this sample. 

\begin{figure}[!htbp]
\begin{center}
\includegraphics[angle=0,width=100mm]{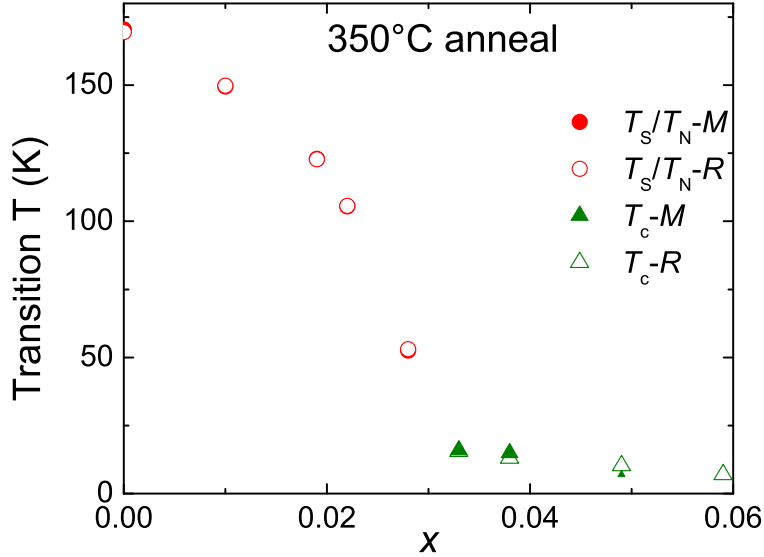}
\end{center}
\caption{(Color online) Phase diagram of transition temperature, {\itshape T}, versus Co concentration, {\itshape x}, of Ca(Fe$_{1-x}$Co$_{x}$)$_{2}$As$_{2}$ samples for an annealing/quenching temperature of 350$^\circ$C. The size of filled triangle ({\itshape T}$_{c}$-{\itshape M}) schematically represents size of low temperature diamagnetic fraction.
The filled symbols are inferred from magnetization ({\itshape M}) data, the open symbols are inferred from resistance ({\itshape R}) data.}
\label{350C-Tx}
\end{figure}

Using the criteria discussed in the Experimental Methods section above, a phase diagram of transition temperature versus Co concentration can be constructed based on the magnetic susceptibility and electric resistance data. Figure \ref{350C-Tx} presents the {\itshape T}-{\itshape x} phase for an annealing/quenching temperature of 350$^\circ$C. The magnetic/structural phase transition is suppressed continuously and the phase line drops to zero for a substitution level between {\itshape x} = 0.028 and {\itshape x} = 0.033, and the superconducting phase emerges by {\itshape x} = 0.033. {\itshape T}$_{c}$ is highest when the antiferromagnetic/orthorhombic phase has just been suppressed completely; {\itshape T}$_{c}$ is suppressed by further Co substitution. The superconducting region extends to around {\itshape x} = 0.059. But, as mentioned above, by this substitution level the superconductivity may just be filamentary. No clear evidence of either the coexistence of the antiferromagnetic/orthorhombic with the superconducting phases or any splitting of the magnetic and structure phase transitions is observed.    

\begin{figure}[!htbp]
\begin{center}
\includegraphics[angle=0,width=160mm]{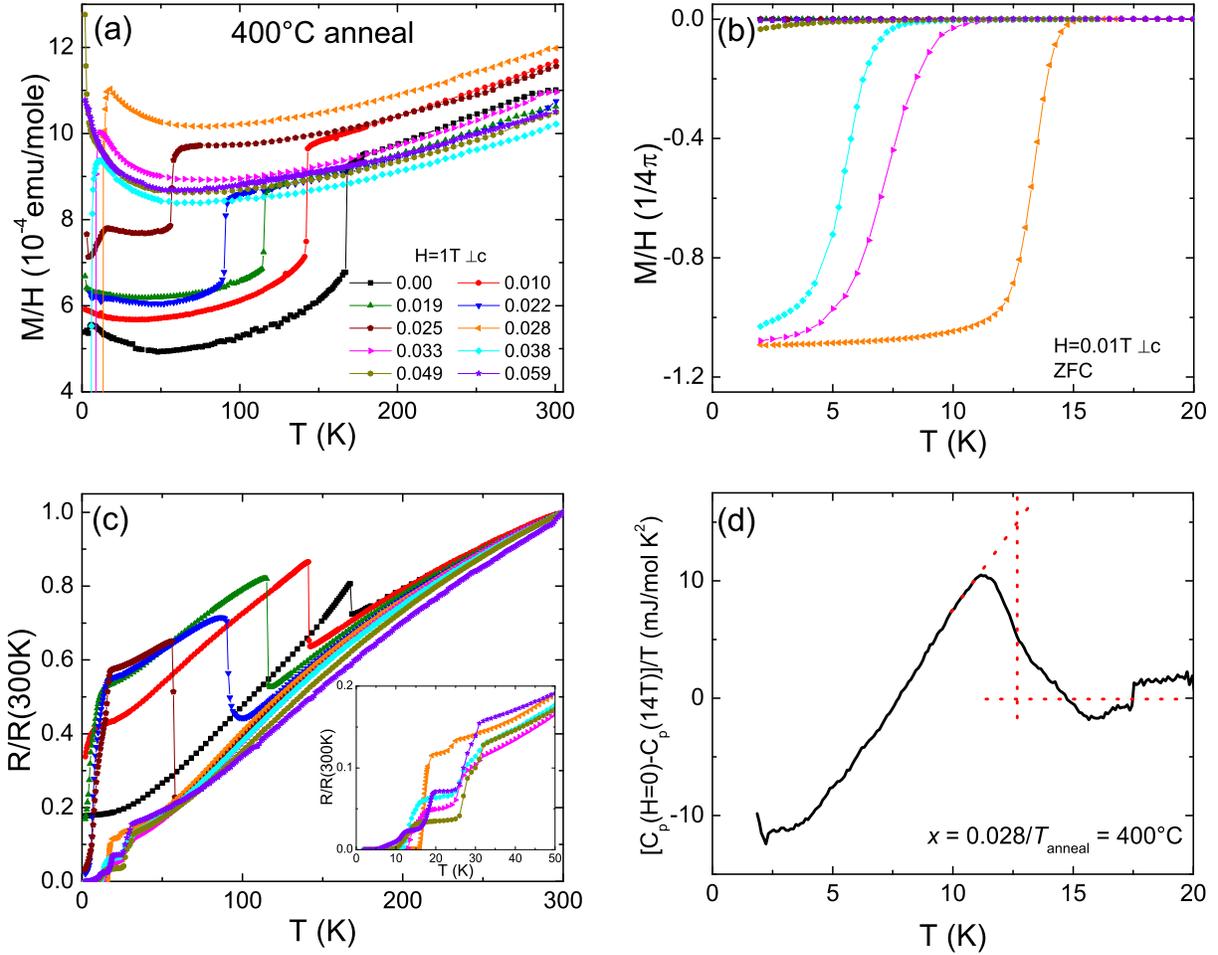}
\end{center}
\caption{(Color online) Temperature dependent (a) magnetic susceptibility with a field of 1 T applied perpendicular to the {\itshape c}-axis, (b) low field magnetic susceptibility measured upon ZFC with a field of 0.01 T applied perpendicular to the {\itshape c}-axis and (c) normalized electrical resistance of Ca(Fe$_{1-x}$Co$_{x}$)$_{2}$As$_{2}$ samples for an annealing/quenching temperature of 400$^\circ$C, together with (d) the specific heat data for the {\itshape x} = 0.028/{\itshape T}$_{anneal}$ = 400$^\circ$C sample (see text). Low temperature resistance of superconducting samples are presented in the inset of (c).}
\label{400CMR}
\end{figure}

\begin{figure}[!htbp]
\begin{center}
\includegraphics[angle=0,width=100mm]{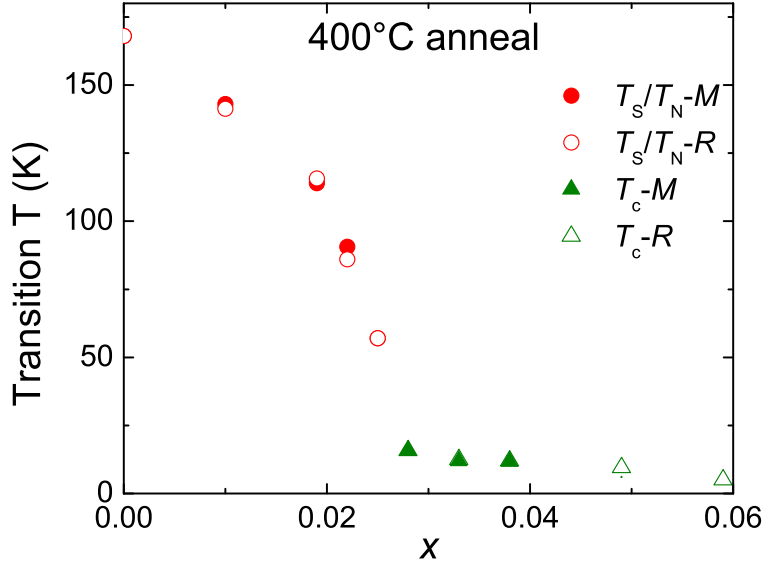}
\end{center}
\caption{(Color online) Phase diagram of transition temperature, {\itshape T}, versus Co concentration, {\itshape x}, of Ca(Fe$_{1-x}$Co$_{x}$)$_{2}$As$_{2}$ samples for an annealing/quenching temperature of 400$^\circ$C. 
The size of filled triangle ({\itshape T}$_{c}$-{\itshape M}) schematically represents size of low temperature diamagnetic fraction.
Filled symbols are inferred from magnetization ({\itshape M}) data, open symbols are inferred from resistance ({\itshape R}) data.}
\label{400C-Tx}
\end{figure}

To further study the effects of the annealing/quenching temperature on this series of compounds, we increased the annealing/quenching temperature to 400$^\circ$C. The magnetic susceptibility and resistance data, as well as specific heat data for the {\itshape x} = 0.028/{\itshape T}$_{anneal}$ = 400$^\circ$C sample, are shown in Fig. \ref{400CMR} and the {\itshape T}-{\itshape x} phase diagram is shown in Fig. \ref{400C-Tx}. As in the case of 350$^\circ$C annealing/quenching, the pure compound is in the antiferromagnetic/orthorhombic state at low temperature. Substituting Co suppresses the antiferromagnetic/structural transition temperature and again, when it is suppressed completely, the superconducting phase appears. The major difference for this higher annealing/quenching temperature is that the {\itshape T}$_{N}$/{\itshape T}$_{S}$ line is suppressed by several K for {\itshape x} = 0 and by {\itshape x} = 0.028, the antiferromagnetic/orthorhombic phase is already suppressed completely and the superconducting phase appears with full diamagnetism whereas, for 350$^\circ$C annealing, this only occurs for {\itshape x} = 0.033. This is consistent with the fact that increasing the annealing/quenching temperature suppresses the antiferromagnetic/structural transition temperature as shown for pure compound in our previous work.\cite{Ran11} The temperature dependent specific heat for {\itshape H} = 0 and {\itshape H} = 14~T for the {\itshape x} = 0.028/{\itshape T}$_{anneal}$ = 400$^\circ$C sample were substracted and the $\Delta${\itshape C}$_{P}$/{\itshape T}$_{c}$ data are consistent with bulk superconductivity (see discussion below). Again neither coexistence of the antiferromagnetic/orthorhombic and the superconducting phases nor splitting of {\itshape T}$_{S}$ and {\itshape T}$_{N}$ were observed. Both {\itshape T}$_{c}$ and diamagnetism fraction are optimal right after the antiferromagnetic/orthorhombic state is completely suppressed and then start to decrease with increasing Co concentration. 

\begin{figure}[!htbp]
\begin{center}
\includegraphics[angle=0,width=160mm]{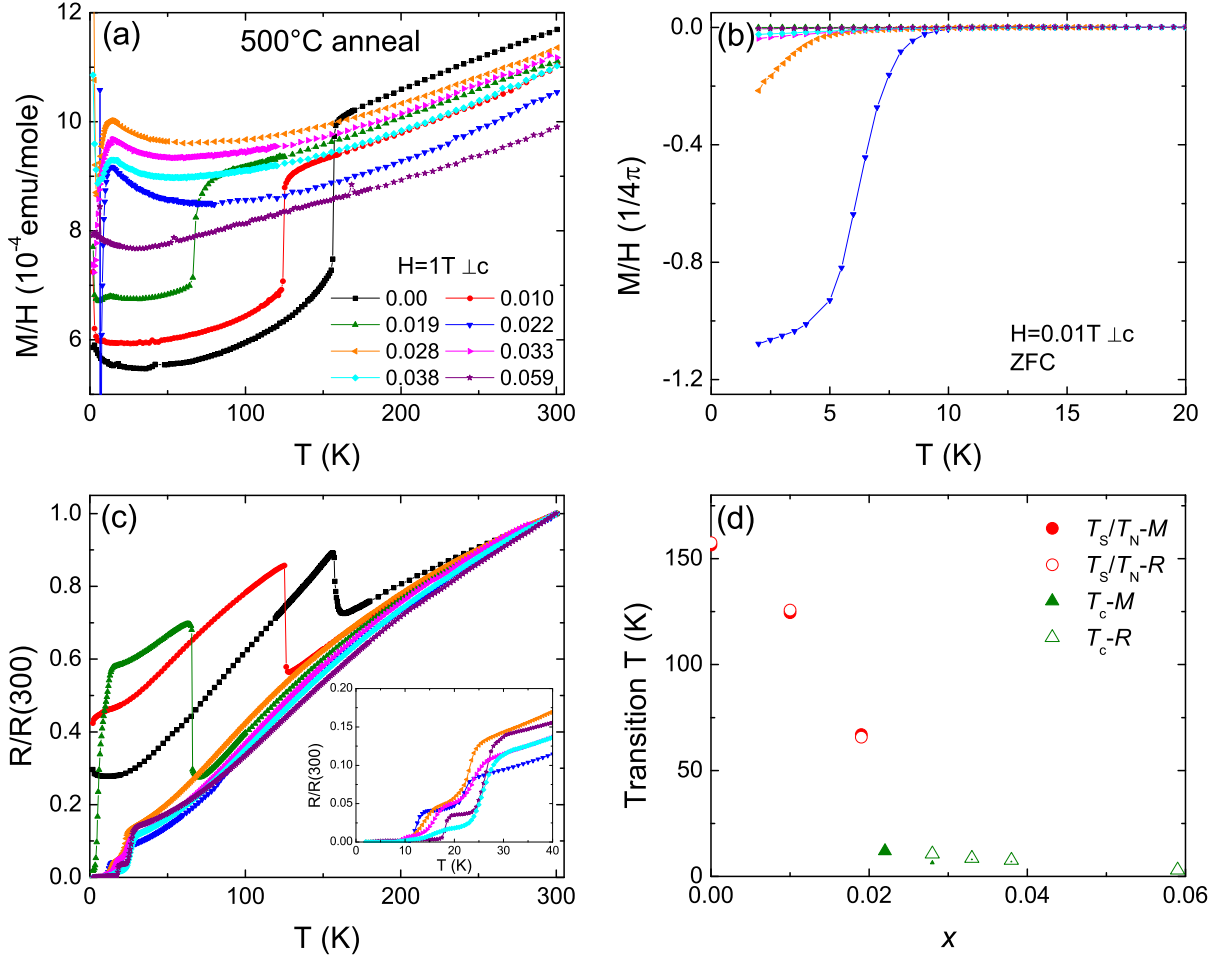}
\end{center}
\caption{(Color online) Temperature dependent (a) magnetic susceptibility with a field of 1 T applied perpendicular to the {\itshape c}-axis, (b) low field magnetic susceptibility measured upon ZFC with a field of 0.01~T applied perpendicular to the {\itshape c}-axis, (c) normalized electrical resistance and (d) phase diagram of transition temperature, {\itshape T}, versus Co concentration, {\itshape x}, of Ca(Fe$_{1-x}$Co$_{x}$)$_{2}$As$_{2}$ samples for an annealing/quenching temperature of 500$^\circ$C. Low temperature resistance of superconducting samples are presented in the inset of (c). In figure (d) the size of filled triangle ({\itshape T}$_{c}$-{\itshape M}) schematically represents size of low temperature diamagnetic fraction. Filled symbols are inferred from magnetization ({\itshape M}) data, open symbols are inferred from resistance ({\itshape R}) data.}
\label{500C-Tx}
\end{figure}

Figure~\ref{500C-Tx} presents the corresponding data for a 500$^\circ$C annealing/quenching temperature. At this annealing/quenching temperature, the antiferromagnetic/structural transition starts with a lower temperature for the pure compound and the switch between the antiferromagnetic/orthorhombic and the superconducting phase occurs between {\itshape x} = 0.019 and 0.022. Only one sample, {\itshape x} = 0.022, shows significant amount of diamagnetism with {\itshape T}$_{c}$ around 9~K.

\begin{figure}[!htbp]
\begin{center}
\includegraphics[angle=0,width=160mm]{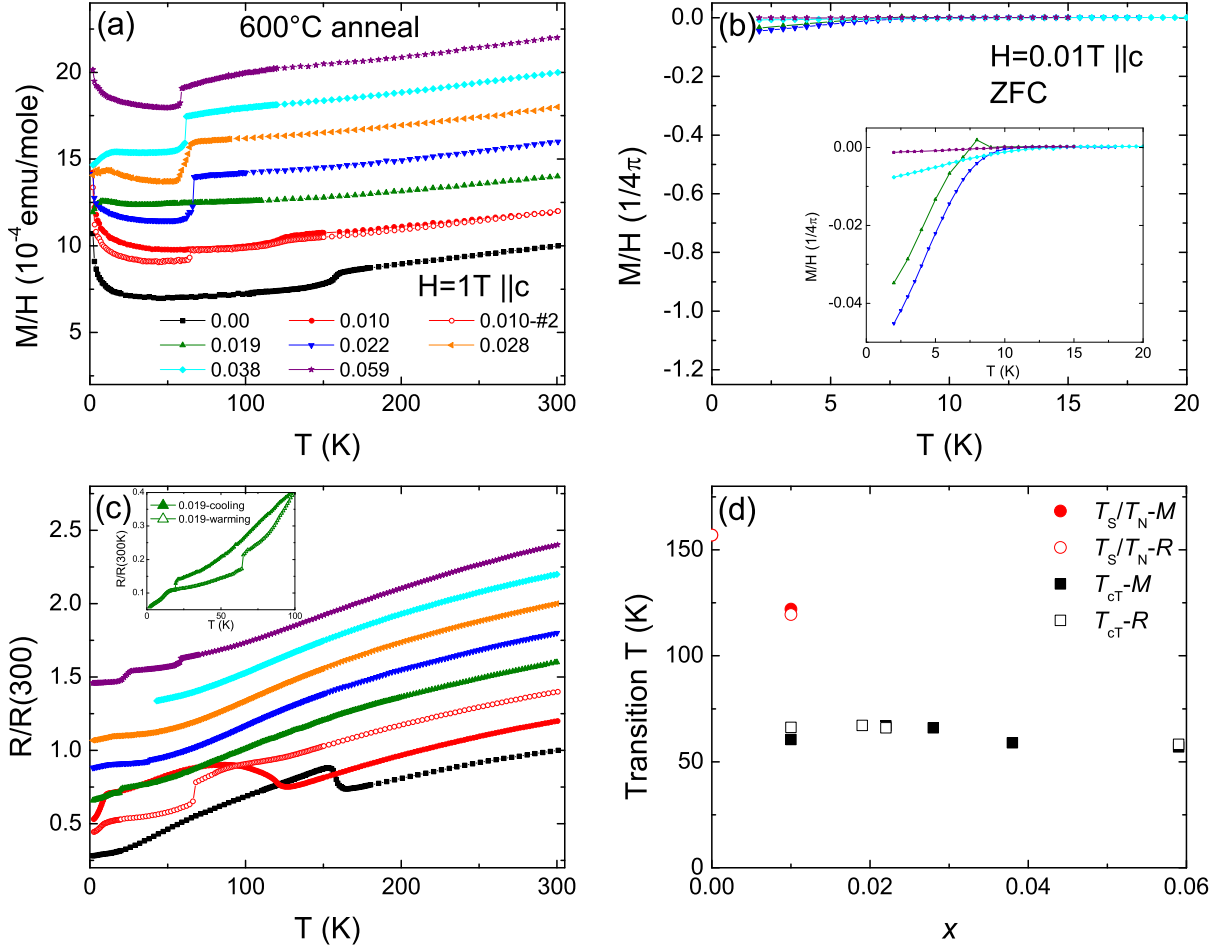}
\end{center}
\caption{(Color online) Temperature dependent (a) magnetic susceptibility with a field of 1 T applied parallel to the {\itshape c}-axis, (b) low field magnetic susceptibility measured upon ZFC with a field of 0.01~T applied parallel to the {\itshape c}-axis, (c) normalized electrical resistance and (d) phase diagram of transition temperature, {\itshape T}, versus Co concentration, {\itshape x}, of Ca(Fe$_{1-x}$Co$_{x}$)$_{2}$As$_{2}$ samples for an annealing/quenching temperature of 600$^\circ$C. The inset of (c) presents the resistance data of 1.9$\%$ sample measured upon warming up and cooling down. For clarity, susceptibility data in (a) have been offset by 2 $\times$ 10$^{-4}$ emu/mole from each other and resistance data in (b) have been offset by 0.2 from each other. In figure (d), the filled symbols are inferred from magnetization ({\itshape M}) data, the open symbols are inferred from resistance ({\itshape R}) data.}
\label{600C-Tx}
\end{figure}

A dramatic change is seen when the annealing/quenching temperature is increased to 600$^\circ$C, as shown in Fig.~\ref{600C-Tx}. The susceptibility is measured with the magnetic field applied parallel to the {\itshape c}-axis, in which direction the size of the jump in susceptibility for the collapsed tetragonal phase transition is significantly larger than that for the antiferromagnetic/structural phase transition, as discussed above, in the Experimental Methods section. Resistance data was also utilized to confirm the nature of the transition since it shows clearly different signature for the two types of phase transition: an upward jump for the antiferromagnetic/structural phase transition and a downward jump or loss of signal for the collapsed tetragonal phase transition. With the combination of these criteria, it can be seen clearly that the pure compound is in the antiferromagnetic/orthorhombic state at low temperature, whereas the samples with {\itshape x} $\textgreater$ 0.022 are in the collapsed tetragonal phase. None of the sample reaches a low-temperature {\itshape R} = 0 state. Figure ~\ref{600C-Tx}c presents the low field susceptibility data. It can be seen, no significant superconducting fraction is observed for sample in either the antiferromagnetic/orthorhombic or the collapsed tetragonal states.

For {\itshape x} = 0.010, two samples were measured. One sample manifests broadened signatures in both susceptibility and resistance that can be associated with the antiferromagnetic/orthorhombic phase transition. The other sample shows double transitions with the upper one consistent with the antiferromagnetic/structural transition and the lower one consistent with the transition into the collapsed tetragonal phase. It is likely that this sample is a mixture of two types of phases, which is reasonable noting that 600$^\circ$C seems to be near the antiferromagnetic/collapsed tetragonal phase boundary and a small degree of inhomogeneity of the local strain may separate the sample into two phases.

For {\itshape x} =0.019, the susceptibility data do not manifest a clear signature of either type of transition whereas resistance measured on the same piece of sample shows a downward jump with hysteresis of $\sim$40K between cooling and warming indicating a transition into the collapsed tetragonal phase, as shown in the inset of Fig. \ref{600C-Tx}c. Given that susceptibility, as a thermodynamic measurement, tells more about the bulk properties, it is possible that only part of the sample is in a collapsed tetragonal state at low temperature. 

Figure~\ref{600C-Tx}d shows the phase diagram for the annealing/quenching temperature of 600$^\circ$C reconstructed from these data. The antiferromagnetic/structural phase transition is suppressed by Co substitution, but unlike the cases of the lower annealing/quenching temperatures, which show a superconducting region when the antiferromagnetic/orthorhombic phase is suppressed completely, here the collapsed tetragonal phase line truncates the suppression of {\itshape T}$_{N}$/{\itshape T}$_{S}$ and no bulk superconducting phase is observed. It is worth noting that although the transition temperature of the antiferromagnetic/orthorhombic phase is suppressed by Co substitution, the transition temperature of the collapsed tetragonal phase stays fairly constant as Co concentration increases. 

\begin{figure}[!htbp]
\begin{center}
\includegraphics[angle=0,width=160mm]{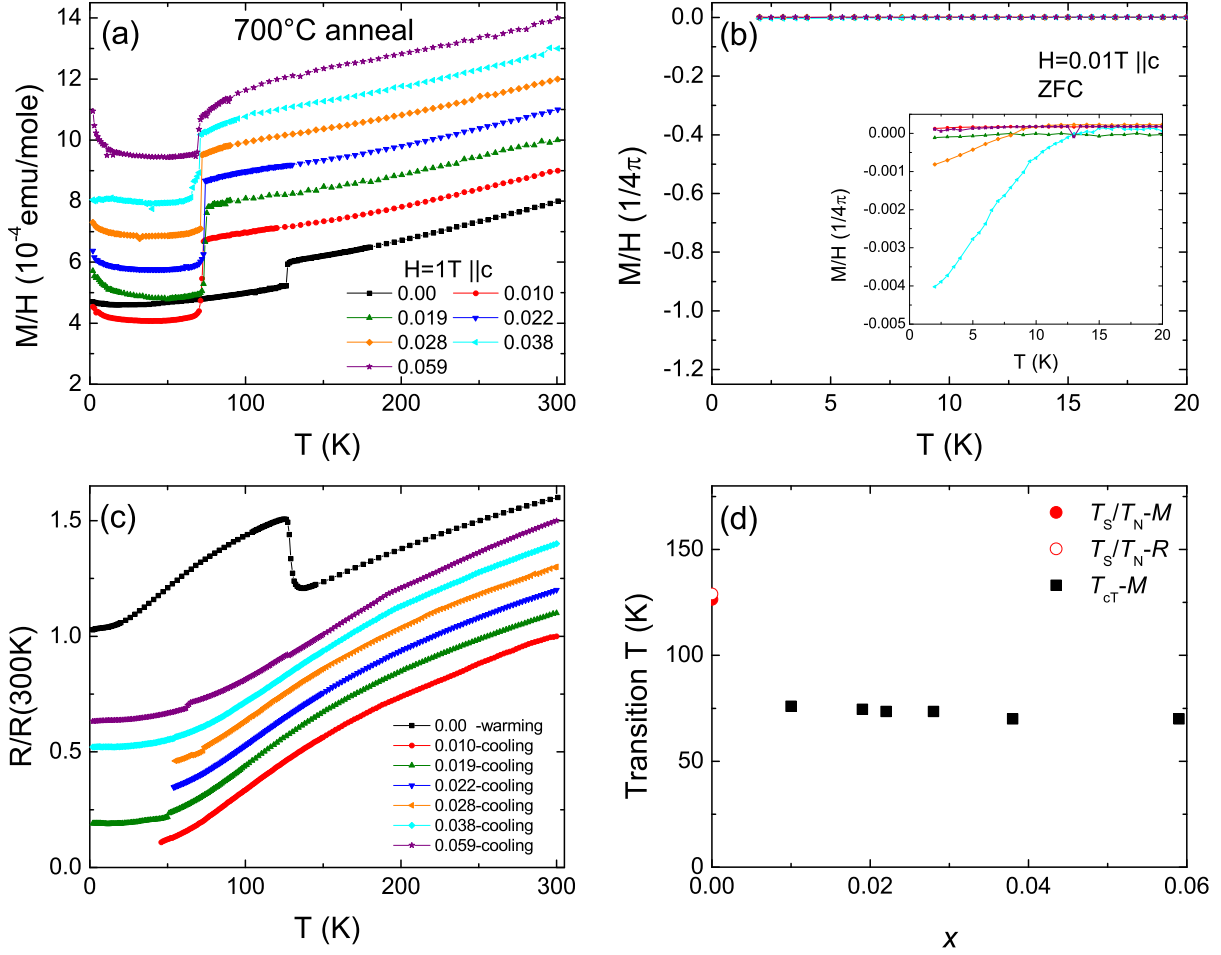}
\end{center}
\caption{(Color online) Temperature dependent (a) magnetic susceptibility with a field of 1 T applied parallel to the {\itshape c}-axis, (b) low field magnetic susceptibility measured upon ZFC with a field of 0.01~T applied parallel to the {\itshape c}-axis, (c) normalized electrical resistance and (d) phase diagram of transition temperature, {\itshape T}, versus Co concentration, {\itshape x}, of Ca(Fe$_{1-x}$Co$_{x}$)$_{2}$As$_{2}$ samples for an annealing/quenching temperature of 700$^\circ$C. For clarity, susceptibility data in (a) have been offset by 1 $\times$ 10$^{-4}$ emu/mole from each other and resistance data in (b) have been offset by 0.1 from each other. In figure (d), the filled symbols are inferred from magnetization ({\itshape M}) data, the open symbols are inferred from resistance ({\itshape R}) data.}
\label{700C-Tx}
\end{figure}

Figures~\ref{700C-Tx}a to \ref{700C-Tx}c present the magnetic susceptibility and normalized resistance data for the annealing/quenching temperature of 700$^\circ$C. Again, the susceptibility is measured with field applied parallel to the {\itshape c}-axis. Both susceptibility and resistance data can be divided into two groups. The signatures in the data from the pure compound clearly show that it's in the antiferromagnetic/orthorhombic state at low temperature. On the other hand, all Co substituted samples show essentially the same signature: very sharp drop in susceptibility and a weak downward jump in resistance which is sometimes accompanied by a loss of contact or continuity due to sample breakage. No significant superconducting fraction is observed, as shown in Fig. \ref{700C-Tx}b. Also {\itshape R}({\itshape T}) data does not show any indication of superconductivity for any substitution level.

Figure~\ref{700C-Tx}d summaries the phase diagram for this annealing/quenching temperature. Similar to the case of the 600$^\circ$C annealing/quenching, the antiferromagnetic/orthorhombic phase only exist when {\itshape T}$_{N}$/{\itshape T}$_{S}$ \textgreater~{\itshape T}$_{cT}$. The transition temperature of collapsed tetragonal state remains roughly constant as Co concentration increases, but {\itshape T}$_{cT}$ is clearly higher for the 700$^\circ$C annealed/quenched samples than it is for the 600$^\circ$C ones, consistent with a continued increase in stress/strain with increasing annealing/quenching temperature.

\begin{figure}[!htbp]
\begin{center}
\includegraphics[angle=0,width=100mm]{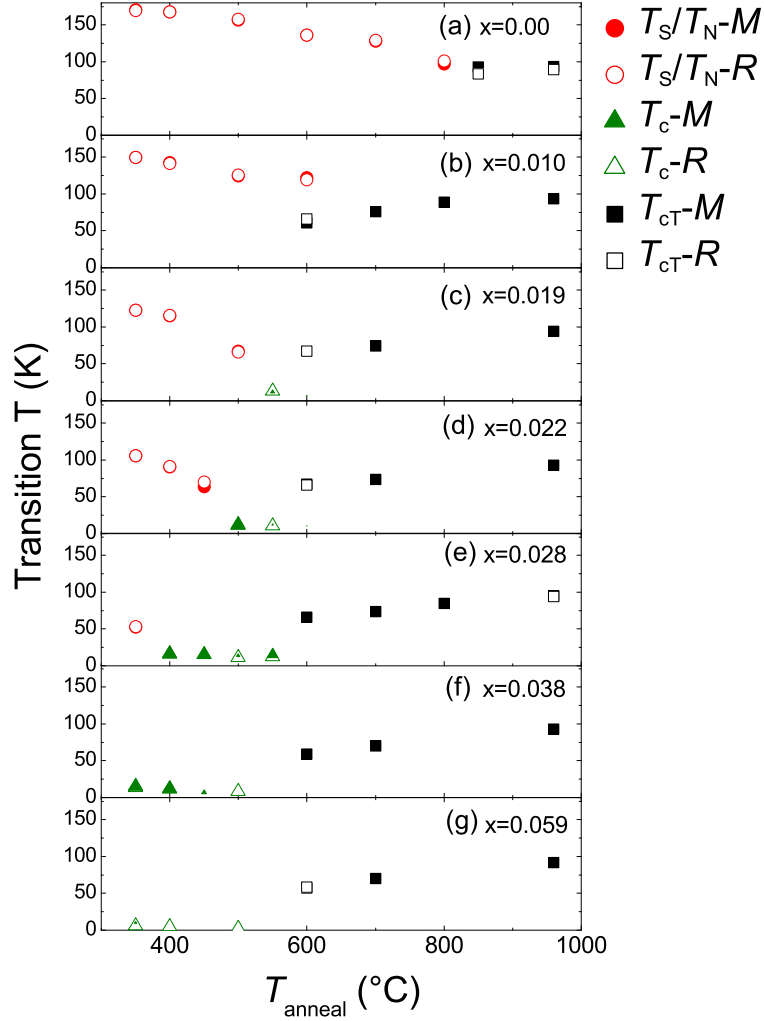}
\end{center}
\caption{(Color online) Phase diagram of transition temperature, {\itshape T}, versus annealing/quenching temperatue, {\itshape T}$_{anneal}$, for \CaCo for (a) {\itshape x} = 0.00, (b) {\itshape x} = 0.010, (c) {\itshape x} = 0.019, (d) {\itshape x} = 0.022, (e) {\itshape x} = 0.028, (d) {\itshape x} = 0.038 and (e) {\itshape x} = 0.059. The size of filled triangle ({\itshape T}$_{c}$-{\itshape M}) schematically represents size of low temperature diamagnetic fraction. Filled symbols are inferred from magnetization ({\itshape M}) data, open symbols are inferred from resistance ({\itshape R}) data.}
\label{TT}
\end{figure}

So far, the phase diagram data have only been shown as {\itshape T}-{\itshape x} cuts for a fixed annealing/quenching temperature. The same set of data can also be presented as phase diagrams of transition temperature versus annealing/quenching temperature ({\itshape T}-{\itshape T}$_{anneal}$ cuts) for each Co substitution level. The {\itshape T}-{\itshape T}$_{anneal}$ phase diagrams are presented in Fig.~\ref{TT}a to \ref{TT}g. For the pure compound,\cite{Ran11} the antiferromagnetic/orthorhombic phase line is suppressed with increasing annealing/quenching temperature and disappears into the collapsed tetragonal phase line at around 800$^\circ$C. No superconductivity is observed. Substituting Co suppresses the antiferromagnetic/orthorhombic phase line. Therefore, for the {\itshape x} = 0.010 sample, the antiferromagnetic/orthorhombic phase line starts at a lower temperature and the entire antiferromagnetic/orthorhombic phase region shrinks. The collapsed tetragonal phase line is further revealed with the antiferromagnetic/orthorhombic phase line merging with it at around 600$^\circ$C, which is a lower annealing/quenching temperature for the onset of the collapsed tetragonal phase than that for the pure compound. For the {\itshape x} = 0.010 sample, the two phase lines still intersect/overlap each other and there is no superconductivity. As the Co concentration is increased further, the antiferromagnetic/orthorhombic phase line is further suppressed but the collapsed tetragonal phase line remains roughly unchanged. There seems to be a minimum of annealing/quenching temperature (internal strain) to stabilize the collapsed tetragonal phase (roughly {\itshape T}$_{anneal}$ = 600$^\circ$C). Therefore, as the antiferromagnetic/orthorhombic phase line is suppressed further, at annealing/quenching temperatures lower than 600$^\circ$C, the two phase lines separate. For {\itshape x} = 0.019, and even more clearly for {\itshape x} = 0.022, the two phase lines no longer intersect each other, leaving an intermediate region where one finds the superconducting phase. Further increasing Co concentration, the antiferromagnetic/orthorhombic phase line is suppressed more and more, and the space between the antiferromagnetic/orthorhombic and the collapsed tetragonal phase lines becomes larger and larger. By {\itshape x} = 0.038, the antiferromagnetic/orthorhombic phase is completely suppressed and the low temperature state is divided into two phases: the superconducting phase and the collapsed tetragonal phase. For {\itshape x} = 0.059, the superconducting signal is rather weak and can only be extracted from resistance data. It is not clear in these cases if any bulk superconductivity remains.

\section{Discussion}

\begin{figure}[!htbp]
\begin{center}
\includegraphics[angle=0,width=100mm]{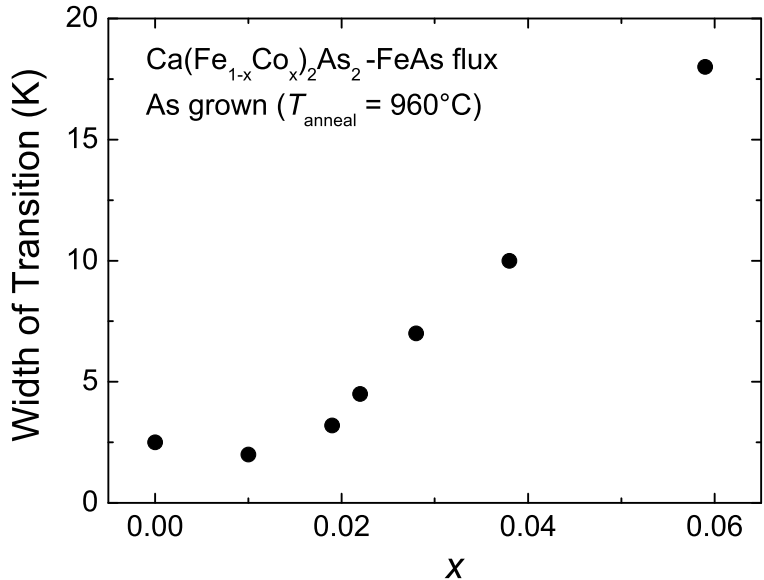}
\end{center}
\caption{width of transition of as-grown Ca(Fe$_{1-x}$Co$_{x}$)$_{2}$As$_{2}$ samples as a function of measured Co concentration, {\itshape x}.}
\label{Asgrown-Tx}
\end{figure}

The thermodynamic, transport and microscopic diffraction measurements of the the {\itshape x} = 0.059, as-grown sample suggest that for the as-grown \CaCo samples there may be a critical end point beyond which the system has a continuous thermal contraction rather than a first order phase transition. Figure~\ref{Asgrown-Tx} presents the width of the transition, which is defined as full width at half maximum of the peak in temperature derivative of magnetic susceptibility. It can be seen that the broadening in transition starts from about {\itshape x} = 0.022. The resistance data shown in Fig.~\ref{Asgrown-MR}b can be divided into two groups according to whether the resistance bar cracks and loses contact when cooling down. It’s clear that the samples with {\itshape x} smaller than 0.028 all lose contacts below the transition temperature indicating these samples undergo first order, structural phase transitions. On the other hand, starting from {\itshape x} = 0.028, the resistance bars survive down to the base temperature of 1.8~K although the resistive data are not ideally smooth. Again these data are consistent with the magnetic susceptibility measurements shown in Fig. \ref{Asgrown-MR}a. To fully address the question of the existence of a critical end point, detailed study of thermodynamic and microscopic properties will be needed, but, at this point the as-grown \CaCo system appears to be a rare example of such isotructural transition  that can be tuned in this manner (the volume collapse in Ce being another such example\cite{Ce}).

\begin{figure}[!htbp]
\begin{center}
\includegraphics[angle=0,width=100mm]{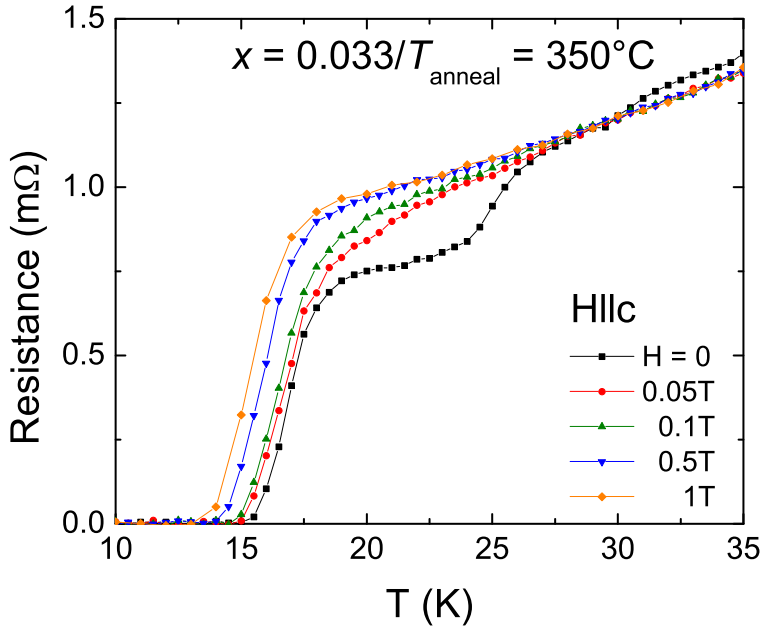}
\end{center}
\caption{(Color online) Temperature dependent resistance data of the {\itshape x} = 0.033/{\itshape T}$_{anneal}$ = 350$^\circ$C sample, measured in zero field and applied field up to 1~T}
\label{Rfield}
\end{figure}

Filamentray superconductivity is a common problem in the AFe$_{2}$As$_{2}$ based materials.\cite{Saha09,Colombier09} In \CaP compounds great care has to be taken to identify and separate filamentray superconductivity from bulk superconductivity. The resistance data show a small superconductivity like drop at around 25~K in many samples before it reaches zero with further cooling. A magnetic field can been applied to these samples and these steps are suppressed by a field as small as 0.05~T. Figure~\ref{Rfield} presents the resistance data, in applied magnetic field, for the {\itshape x} = 0.033/{\itshape T}$_{anneal}$ = 350$^\circ$C sample, as an example. In a field of 0.05~T, the drop at higher temperature is suppressed completely whereas the final step towards zero remains sharp and is only slightly shifted to lower temperature. This indicates the final step is a rather robust signature of superconductivity, although the question of why the 25~K feature (whatever its origin is) has such an extreme field dependence is left as an unsolved puzzle for now.

\begin{figure}[!htbp]
\begin{center}
\includegraphics[angle=0,width=100mm]{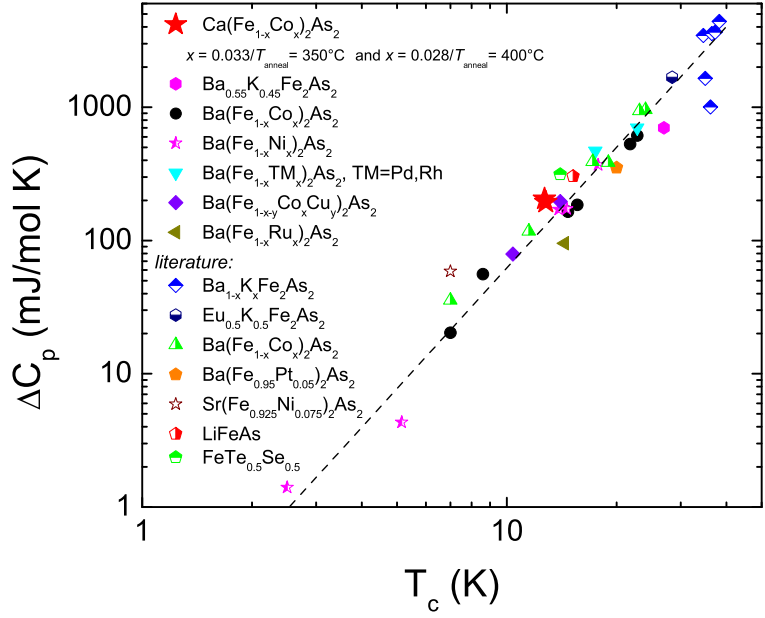}
\end{center}
\caption{(Color online) $\Delta${\itshape C}$_{P}$ vs {\itshape T}$_{c}$ for the {\itshape x} = 0.033/{\itshape T}$_{anneal}$ = 350$^\circ$C sample and the {\itshape x} = 0.028/{\itshape T}$_{anneal}$ = 400$^\circ$C sample, plotted together with literature data for various FeAs-based superconducting materials.\cite{Sergey09cp,Kogan09,Matsuda11}}
\label{BNC}
\end{figure}

In order to confirm bulk superconductivity, thermodynamic measurements are needed. Whereas low field magnetization data can be suggestive, specific heat data is even clearer evidence. Specific heat measurements were made on the representative samples, the {\itshape x} = 0.033/{\itshape T}$_{anneal}$ = 350$^\circ$C sample (Fig.~\ref{3.3-350C-Cp}) and the {\itshape x} = 0.028/{\itshape T}$_{anneal}$ = 400$^\circ$C sample (Fig.~\ref{400CMR}), both of which are located in close approximity to the suppressed {\itshape T}$_{N}$/{\itshape T}$_{S}$ line and both of which show full diamagnetic fraction in zero field cooling. $\Delta${\itshape C}$_{P}$/{\itshape T}$_{c}$ values of 16.1~mJ/mol~K$^2$ and 15.1~mJ/mol~K$^2$ are inferred from the data for the the {\itshape x} = 0.033 and the {\itshape x} = 0.028 samples, respectively. These values can be placed in context of other substituted AFe$_{2}$As$_{2}$ compounds on a plot of $\Delta${\itshape C}$_{P}$({\itshape T}$_{c}$) (Fig. \ref{BNC}).\cite{Stewart11,Sergey09cp,Kogan09,Matsuda11} Based on this comparison we can see that the signature of superconductivity found in specific heat data from these samples is comparable to that of Ba122 with various substitutions and other iron-based superconducting compounds. This is in contrast to the previously reported rare earth substituted Ca122, in which case no clear evidence of bulk superconductivity is observed.\cite{Saha12}

\begin{figure}[!htbp]
\begin{center}
\includegraphics[angle=0,width=100mm]{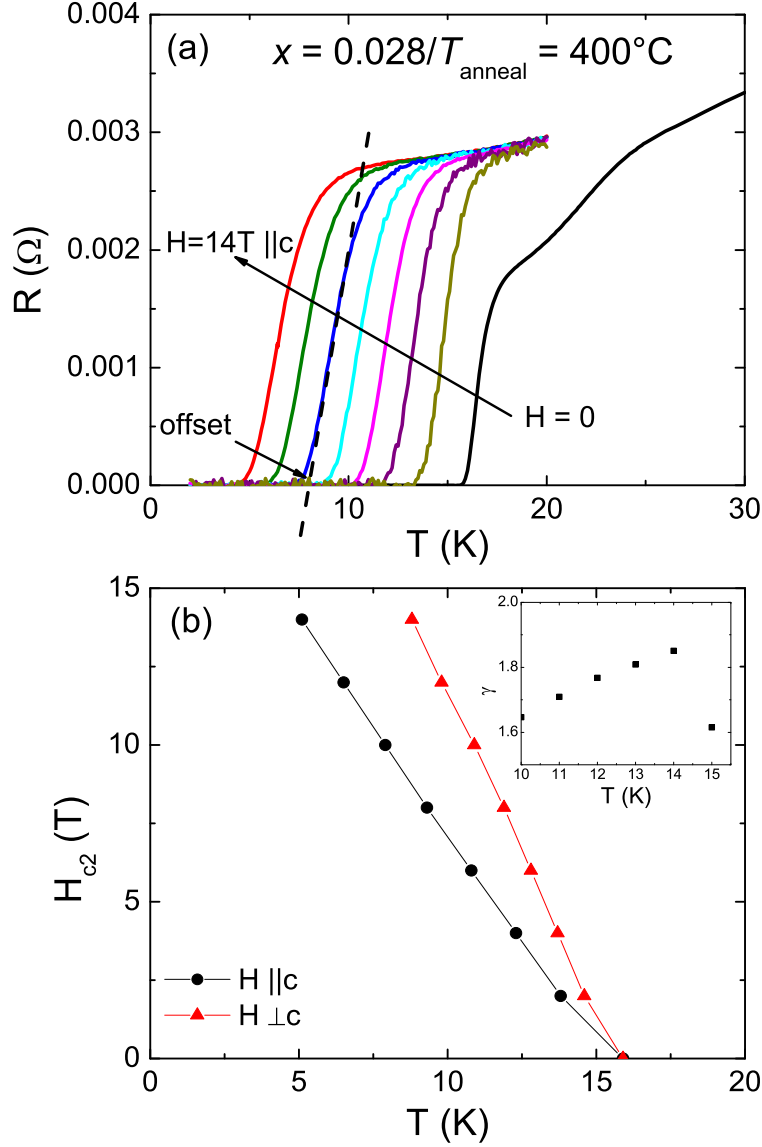}
\end{center}
\caption{(Color online) Temperature dependent (a) resistance data of the {\itshape x} = 0.028/{\itshape T}$_{anneal}$ = 400$^\circ$C sample, measured in  applied field parallel to the {\itshape c}-axis for {\itshape H} = 0, 2~T, 4~T, 6~T, 8~T, 10~T, 12~T and 14~T and (b) anisotropic {\itshape H}$_{c2}$ data determined from {\itshape R}({\itshape T}) data. Inset to (b) shows $\gamma$ = {\itshape H}$_{c2}^{\perp c}$/{\itshape H}$_{c2}^{\parallel c}$ for 10~K $\textless$ {\itshape T} $\textless$ 16~K.}
\label{Hc2}
\end{figure}

To more fully characterize the superconducting state, temperature dependent anisotropic {\itshape H}$_{c2}$ was measured on the {\itshape x} = 0.028/{\itshape T}$_{anneal}$ = 400$^\circ$C sample up to 14~T. The {\itshape R}({\itshape T}) data for various {\itshape H} in the direction parallel to the {\itshape c}-axis are shown in Fig. \ref{Hc2}a along with an example of the criterion used to infer {\itshape H}$_{c2}$, offset of the superconducting transition. Figure~\ref{Hc2}b presents the anisotropic {\itshape H}$_{c2}$ plot inferred from the {\itshape R}({\itshape T}) data and, in the inset, the temperature dependence of $\gamma$ = {\itshape H}$_{c2}^{\perp c}$/{\itshape H}$_{c2}^{\parallel c}$. After an initial upward curvature, there is roughly a linear increase of {\itshape H}$_{c2}$ with decreasing temperature. {\itshape H}$_{c2}$ at zero temperature, although is not reached in our measurement, seems to be $\sim$20~T. As can be seen in the inset of Fig. \ref{Hc2}b, the $\gamma$ has values between 1.5 and 2.0. These values are consistent with those found for K-substituted, Co-substituted and Ni-substituted Ba122 samples.\cite{Altarawneh08,Ni08BaCo,Ni10TM}

The progression of the {\itshape T}-{\itshape T}$_{anneal}$ phase diagrams (Fig.~\ref{TT}) from the pure compound to the highest substitution level reveals that there is no coexistence of superconductivity with either the antiferromagnetic/orthorhombic phase or the collapsed tetragonal phase. The absence of the superconductivity in the collapsed tetragonal phase region is consistent with the idea that the mechanism of iron-based superconductor depends on magnetic fluctuations. Since in the collapsed tetragonal phase magnetic moment is quenched completely, there is no spin fluctuation to drive the superconducting phase.\cite{Pratt09}

\begin{figure}[!htbp]
\begin{center}
\includegraphics[angle=0,width=100mm]{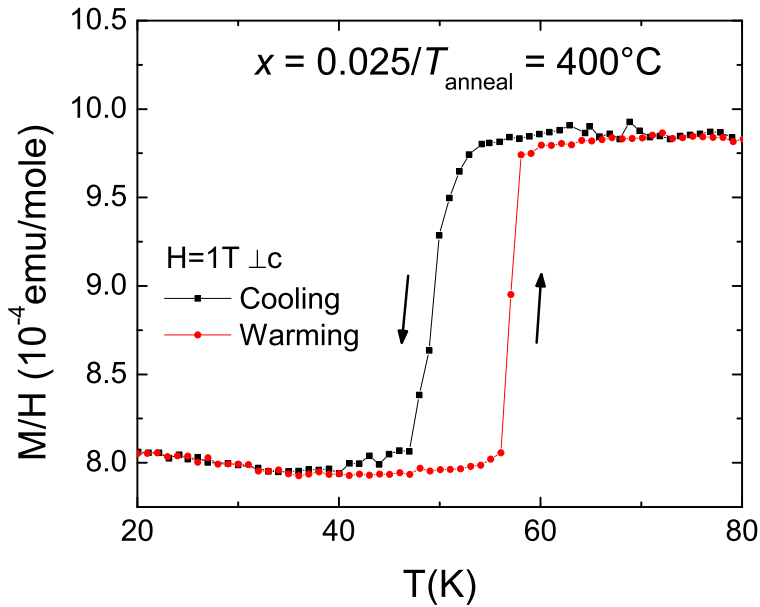}
\end{center}
\caption{(Color online) Temperature dependent magnetic susceptibility of the {\itshape x} = 0.025/{\itshape T}$_{anneal}$ = 400$^\circ$C sample, measured upon warming up and cooling down.}
\label{2.5hysteresis}
\end{figure}

The absence of superconductivity in the antiferromagnetic/orthorhombic phase region can be understood based on the fact that the antiferromagnetic/structural phase transition remains quite first order even when it is suppressed to around 50~K, which is the lowest {\itshape T}$_{N}$/{\itshape T}$_{S}$ we obtained in these studies. The first order nature of the antiferromagnetic/structural phase transition is demonstrated by the sharpness of both the magnetic and resistive signatures of the transition as well as the hysteresis of the transition temperature of about 7~K, e.g. the susceptibility data of the {\itshape x} = 0.025/{\itshape T}$_{anneal}$ = 400$^\circ$C sample are shown in Fig. \ref{2.5hysteresis}. The strongly first order nature of the magnetic/structural phase transition in \CaCo is in stark contrast to Ba(Fe$_{1-x}$Co$_{x}$)$_{2}$As$_{2}$ which manifest split, second order magnetic and structural phase transitions.\cite{JPSJFeAs,Hosono09,Chu09,Prozorov10,Canfield10,Johnston10,Stewart11,Ni11review} For small Co substitution levels, in the case of Ba(Fe$_{1-x}$Co$_{x}$)$_{2}$As$_{2}$, a coexisting superconducting state emerges under the suppressed and separated second order phase transitions whereas for \CaCo the superconducting state does not emerge anywhere below the strongly first order, coupled magnetic/structural transition line. This clear difference is also consistent with magnetic fluctuations being vital for the emergence of the superconducting state.

\begin{figure}[!htbp]
\begin{center}
\includegraphics[angle=0,width=160mm]{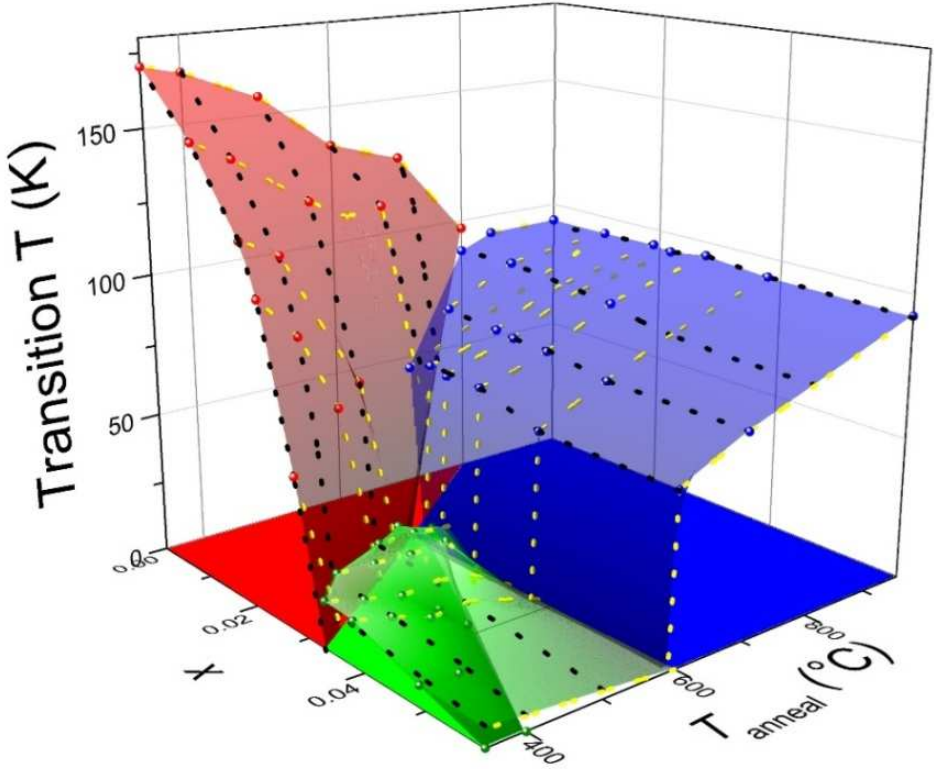}
\end{center}
\caption{(Color online) Three dimensional phase diagram with substitution level, {\itshape x}, annealing/quenching temperature, {\itshape T}$_{anneal}$, and transition temperature, {\itshape T}, as three axes. Red (antiferromagnetic/orthorhombic), green (superconducting) and blue (collapsed tetragonal) spheres represent data. Transparent, colored surfaces are guides to the eyes. Black dashed lines are {\itshape T}-{\itshape x} lines for different {\itshape T}$_{anneal}$ and yellow dashed lines are {\itshape T}-{\itshape T}$_{anneal}$ lines for different {\itshape x}. Solid, colored areas on the {\itshape T}$_{anneal}$-{\itshape x} plane are low temperature ground states. Note: dark green area (from magnetization data) represents the bulk superconducting region whereas light green area (from resistance data) represents the zero resistance region.}
\label{3D}
\end{figure}

With annealing/quenching temperature as another tuning parameter, the phase diagram is essentially extended from two dimensions to three dimensions. We can establish a three dimensional phase diagram with substitution level, {\itshape x}, annealing/quenching temperature, {\itshape T}$_{anneal}$, and transition temperature, {\itshape T}, as the three axes, as shown in Fig.~\ref{3D}. Whereas the antiferromagnetic/orthorhombic phase is clearly suppressed by increasing {\itshape x} and {\itshape T}$_{anneal}$, the collapsed tetragonal phase, once it emerges, varies with {\itshape T}$_{anneal}$, but over this limited substitution range, does not vary significantly with {\itshape x}.  At lowest temperatures there is no co-existence between any of these phases with superconductivity being truncated at low {\itshape x} and low {\itshape T}$_{anneal}$ by the antiferromagnetic/orthorhombic phase and at high {\itshape T}$_{anneal}$ by the collapsed tetragonal phase.

We can compare this 3D phase diagram to that of the rare earth substituted Ca122 system, which can be considered as a combination of electron substitution and chemical pressure.\cite{Saha12} Since it is not clear whether the superconductivity observed in rare earth substituted samples is bulk, we focus on the magnetic/structural phase transition region as well as the collapsed tetragonal phase region. The basic structure of the phase diagrams looks similar. Both substitution and effective pressure (in case of Co substitution, it is annealing and in case of RE substitution, it is chemical pressure) suppresses the antiferromagnetic/orthorhombic phase. The rate of suppression, when calculated in terms of extra electrons, is much higher for Co substitution. When annealed/quenched at 350$^\circ$C, the annealing/quenching temperature at which stress and strain is released to the largest extend in our study, by 3.3$\%$ Co the antiferromagnetic/orthorhombic phase is suppressed completely. Whereas for La substitution, which does not cause a significant change of {\itshape c}-lattice parameter, the antiferromagnetic/orthorhombic phase still survives for 10$\%$ La.\cite{Saha12} This difference can be understood based on the assumption that rare earth substitution perturbs the Ca layer whereas Co substitution perturbs Fe layer. Similarly, in the case of Ba122, K substitution, which perturbs Ba layer, suppresses the antiferromagnetic/orthorhombic phase at a much slower rate than Co substitution, which perturbs Fe layer.\cite{Ni08BaK,Johrendt09,Ni08BaCo}

The three dimensional {\itshape T}-{\itshape x}-{\itshape T}$_{anneal}$ phase diagram we find for Co substitution can also be compared to the earlier data measured on Co substituted samples grown out of Sn\cite{Harnagea11,Matusiak10,Hu11Ca} as well as some recent results on \textquotedblleft furace cooled$\textquotedblright$ Rh substituted samples grown out of FeAs\cite{Danura11}. For the Sn grown \CaCo compounds, our low annealing/quenching temperature ({\itshape T}$_{anneal}$ = 350$^\circ$C) data is qualitatively similar in that there is a suppression of the magnetic/structural phase transition and the appearance of superconductivity. Quantitatively, we find a slightly more rapid suppression of the {\itshape T}$_{N}$/{\itshape T}$_{S}$ line, and a much clearer and systematic evolution of the first order signatures of the magnetic/structural and the collapsed tetragonal phase transitions with substitution and annealing/quenching temperature. Recent Rh substitution work on samples that were allowed to cool to room temperature after a slow cool between 1100 and 1050$^\circ$C found that a very narrow region of Rh substitution revealed partial screening in magnetic susceptibility data for {\itshape x} $\sim$0.02 between the suppression of the {\itshape T}$_{N}$/{\itshape T}$_{S}$ line for low substitution levels and a rapid increase in the collapsed tetragonal phase transition temperature for {\itshape x} greater than 0.024.  It is very likely that a systematic study of the effects of annealing/quenching temperature on FeAs grown, Rh substituted samples will reveal a comprehensible, three dimensional phase diagram, perhaps different from Co substitution in some details due to the different effect of Rh and Co on the size of the {\itshape c}-lattice parameter.

Fianlly, we would like to point out that controlled annealing/quenching of FeAs grown \CaCo opens up a myriad of opportunities for the further research. We are able to tune the system systematically and reproducibly. Given the similar effects of pressure and annealing/quenching temperature, it is now possible for APRES and/or STM to explore what were inaccessible {\itshape T}-{\itshape P} phase diagrams via use of the {\itshape T}-{\itshape x}-{\itshape T}$_{anneal}$ phase diagram. Furthermore, if we extend the {\itshape P}-{\itshape T}$_{anneal}$ analogy from our annealing work on the pure compound\cite{Ran11}, then we expect that continuous tuning can be achieved for Co substituted samples with hydrostatic pressure using He gas medium. For example, the {\itshape T}-{\itshape T}$_{anneal}$ phase diagram of {\itshape x} = 0.022 sample presented in Fig. \ref{TT}d suggests that it might be possible to tuning the system from the antiferromagnetic/orthorhombic phase to the superconducting phase and then to the collapsed tetragonal phase with applied pressures of less than 0.5~GPa. If this is the case, then elastic and inelastic neutron scattering studies on a single sample can be used  to systematically study the magnetic order and fluctuations across the whole phase space of FeAs-based superconductivity.

\section{Conclusions}

We have grown single crystal samples of Co substituted \CaP out of an FeAs flux and found that the as-grown samples are still in the collapsed tetragonal state at low temperature at ambient pressure, similar to the pure compound\cite{Ran11}. We systematically studied effects of annealing/quenching temperatures on the physical properties of these samples. The progression of the {\itshape T}-{\itshape T}$_{anneal}$ phase diagram with increasing Co concentration shows that by substituting Co, the antiferromagnetic/orthorhombic and the collapsed tetragonal phase lines are separated and bulk superconductivity is revealed. We established a 3D phase diagram with Co concentration and annealing/quenching temperature as two independent control parameters. At 2~K the superconducting state exists between a low {\itshape x}, low {\itshape T}$_{anneal}$, antiferromagnetic/orthorhombic phase and a high {\itshape T}$_{anneal}$, collapsed tetragonal phase, in a region where magnetic fluctuations can persist to low enough temperatures.

\section{Acknowledgement}
The author thanks V. Taufour and T. Li for help with figures and N. Ni (as well as F. Fe and C. Co) for useful discussion. Work at the Ames Laboratory was supported by the Department of Energy, Basic Energy Sciences, Division of Materials Sciences and Engineering under Contract No. DE-AC02-07CH11358. S.L.B. acknowledges partial support from the State of Iowa through Iowa State University.


\label{lastpage}

\end{document}